\documentclass[11pt]{article}

\usepackage{graphicx}
\usepackage{endfloat}
\usepackage{amssymb}
\usepackage{slashbox}

\RequirePackage[OT1]{fontenc}
\RequirePackage{amsmath}
\RequirePackage[colorlinks,citecolor=blue,urlcolor=blue]{hyperref}
\RequirePackage{hypernat}
\usepackage[ruled]{algorithm2e}
\usepackage{algorithmic}
\usepackage{graphicx,psfrag,epsf}
\usepackage{epstopdf}
\usepackage{subfloat}
\usepackage{rotating}
\usepackage{subfigure}
\usepackage{longtable}
\usepackage{engord}
\usepackage{url}
\usepackage{enumerate}
\usepackage{color}
\usepackage{layout}
\usepackage{theorem}
\usepackage{amssymb}
\usepackage{float}
\usepackage{hhline}
\usepackage{version}
\usepackage{fancyheadings}
\usepackage{appendix}
\usepackage{epstopdf}

\usepackage[default]{jasa_harvard}
\usepackage{JASA_manu}

\newcommand {\ctn}{\citeasnoun} % change to \citet if using natbib
\newcommand {\ctp}{\cite}       % change to \citep if using natbib
\newtheorem{defi}{Definition}

\newtheorem{thm}{Theorem}%[section]
\newtheorem{pf}{Proof}

\DeclareMathOperator*{\argmin}{\rm argmin}

\pdfoutput=1

\begin{document}

\title{Estimating the Number of Sources in Magnetoencephalography Using Spiked Population Eigenvalues}

\author{Zhigang Yao\\
Department of Statistics and Applied Probability  \\ 21 Lower Kent Ridge Road\\
National University of Singapore, Singapore 117546  \\
\vspace{0.15 in}
email: \texttt{zhigang.yao@nus.edu.sg}\\
Ye Zhang\\
School of Science and Technology\\
\"{O}rebro University \\  SE-701 82  \"{O}rebro, Sweden\\
\vspace{0.15 in}
email: \texttt{ye.zhang@oru.se}\\
Zhidong Bai\\
KLASMOE and School of Mathematics and Statistics\\
Northeast Normal University\\  130024 Changchun, China\\
\vspace{0.15 in}
email: \texttt{baizd@nenu.edu.cn}\\
William F. Eddy\\
Department of Statistics\\
Carnegie Mellon University \\ Pittsburgh, Pennsylvania 15213\\
email: \texttt{bill@stat.cmu.edu}}

\maketitle

\newpage

\mbox{}
\vspace*{2in}
\begin{center}
\textbf{Authors' Footnote:}
\end{center}
Zhigang Yao is Assistant Professor, Department of Statistics and Applied Probability, National University of Singapore, Singapore. Ye Zhang is Research Fellow, School of Science and Technology, \"{O}rebro University, Sweden. Zhidong Bai is Professor, KLASMOE and School of Mathematics and Statistics, Northeast Normal University, China. William F. Eddy is Professor, Department of Statistics, Carnegie Mellon University. Research supported by Yao's MOE Tier 1 and Tier 2 Grant Award. We thank Dr. Anto Bagic for collecting the empty room MEG data and Professor Rob Kass for the BCI data. Our warm thanks also go to three referees and the associate editor for a number of constructive comments.
\newpage
\begin{center}
\textbf{Abstract}
\end{center}
Magnetoencephalography (MEG) is an advanced imaging technique used to measure the magnetic fields outside the human head produced by the electrical activity inside the brain. Various source localization methods in MEG require the knowledge of the underlying active sources, which are identified by {\it a priori}. Common methods used to estimate the number of sources include principal component analysis or information criterion methods, both of which make use of the eigenvalue distribution of the data, thus avoiding solving the time-consuming inverse problem. Unfortunately, all these methods are very sensitive to the signal-to-noise ratio (SNR), as examining the sample extreme eigenvalues does not necessarily reflect the perturbation of the population ones. To uncover the unknown sources from the very noisy MEG data, we introduce a framework, referred to as the intrinsic dimensionality (ID) of the optimal transformation for the SNR rescaling functional. It is defined as the number of the spiked population eigenvalues of the associated transformed data matrix. It is shown that the ID yields a more reasonable estimate for the number of sources than its sample counterparts, especially when the SNR is small. By means of examples, we illustrate that the new method is able to capture the number of signal sources in MEG that can escape  PCA or other information criterion based methods.

\vspace*{.3in}


\newpage

\section{Introduction}
\label{intro}

Thousands of synchronized neurons give rise to macroscopic oscillations, which can be observed in the electroencephalogram (EEG). Meanwhile, the electric currents generated by those synchronized neurons induce extremely weak magnetic fields ($10-100$ femto-Tesla). Measuring magnetic fields of this magnitude is a great challenge. The recent development of superconducting quantum interference devices (SQUIDs) makes it possible to detect these extremely weak magnetic signals. Magnetoencephalography (MEG) is a non-invasive imaging technique that measures the magnetic fields produced by neuronal activity in the brain, with many coils placed around the head \ctp{Cohen_1968}. The fundamental difference between MEG and other functional imaging modalities, such as positron emission tomography (PET) or functional magnetic resonance imaging (fMRI), is that the neuronal activity is measured directly by MEG, in the sense that the signal it measures is a direct consequence of brain activity, but indirectly by PET or fMRI, which measure the blood flow changes or other vascular phenomena. Because of its impressive temporal resolution (better than 1 millisecond, compared to 1 second for fMRI, or up to 1 minute for PET), MEG measures brain activity without any time delay, and this makes MEG a near optimal tool for studying the brain.

In MEG, the sources are mainly the electric currents generated by the activated neurons in the human cortex, and they are usually formulated as mathematical point current dipoles. A brief description of the mathematical modelling and the corresponding forward and inverse problems in MEG can be found in the supplementary materials. The main challenge posed by MEG is to determine the location of electrical activity within the brain from the induced magnetic fields around the skull. This is a typical ill-posed problem since (i) one can always construct a large number of nontrivial dipoles that have a vanishing magnetic field outside the head. (ii) the identification process is usually unstable, i.e., a small amount of noise in the measurement data can lead to enormous errors in the estimates. To tackle the ill-posedness of the inverse MEG problem, a sequence of regularization algorithms have been extensively exploited during the past two decades: see \ctn{Hamalainen_1994}, \ctn{Uutela_1999}, \ctn{Urbano_2000}, \ctn{Pulvermuller_2003}, \ctn**{Mattouta_2006} and references therein. However, almost all of them are highly restricted by the pre-defined number of dipoles or the prior distribution assumed in the data \ctp{Schmidt_1999} . A fundamental challenge here is that it is almost impossible to pick up how many active dipoles will be needed in advance. Other approaches, also known as ``imaging methods'', represented by the multiple signal classification (MUSIC) (see e.g., \ctn{Mosher_1998}), attempt to estimate the number of dipoles (less than $10$) by separating the signal subspace and the noise subspace; that is, they decompose the eigenvalues of the sample space of the MEG data into the eigenvalues of the signal subspace and noise subspace, following the assumption that the lead field vector at each source location is orthogonal to the noise.  By assuming all dipoles are uncorrelated, the beamformer-inspired approaches \ctp{Van_Veen1992} solve a more ``tangent'' problem; that is, avoiding the need to estimate the {\it prior}. Beamformers are essentially spatial filters that suppress the dipoles in a number of selected locations, blocking out the signal originated elsewhere. Most recently, several works based on time-varying dipoles \cite**{Yao1} have been proposed, where it is suggested that the number of varying dipoles is estimated in a dynamic fashion.

%%%%%%%%%%%%%%%%%%%%%%%%%%%%%%%%%%%%%%%%%%%%%

Though the underlying number of dipoles is generally unknown and thus impossible to verify, there has been increasing demand for a reliable estimation. Due to the nature of the inverse problem, estimating the number of dipoles in MEG can pose unconventional challenges.
%Natural as an additive signal-measurement model may be, described in the general framework (\ref{Y_t}), %that is, the data is considered as the sum of noise and a vector of certain linear combinations of random functions related to the dipoles,
The same type of problem can be traced back to the problem of detecting the number of signal sources in the presence of noise, or its {\it intrinsic dimensionality}, in a multiple channels of time series \cite**{Ligget1973,Schmidt_1986}.  The heuristics are as follows: if the noise is Gaussian, then the number of the dipoles is related to the multiplicity of the smallest eigenvalues of the sample covariance matrix, and in this case the key to estimating the number of dipoles is to find a threshold that separates the {\it spiked} eigenvalues, the extreme eigenvalues,  from the {\it bulk} eigenvalues of sample covariance matrix, or, more generally, to perform a sequence of hypothesis tests on the same eigenvalues against the threshold \cite**{Bartlett1954,Lawley956}. Centering around making use of the distribution of eigenvalues of the sample covariance matrix, there are two schools of thoughts in the literature: 1) pursuing the number of signal sources from principal component analysis (PCA), %\ctp{Richards_1993},
independent component analysis (ICA) \cite**{Green_2001} and factor analysis \cite{Malinowski_1977_1,Malinowski_1977_2}; 2) estimating the solution based on certain information criterion, such as Akaike information criterion (AIC) and minimum description length (MDL). This being said, separating the spiked eigenvalues from the {\it sample} covariance does not necessarily give an accurate estimation for the number of sources. The reasons are two-fold: 1) the sample eigenvalues are not consistent estimators of their population counterparts, and hence any test based on sample eigenvalues inevitably reveals considerable deviation from the truth, and 2) the unknown noise structure, which, although it can be estimated, amplifies the difficulty of estimating the transition function (to be defined in Section 2) between the sample eigenvalues and the population eigenvalues. The first one is well illustrated by considering an extreme case (see \ctn{bai1999} for details): assuming the population covariance $\sigma^2 I_{p}$ with finite fourth moment and $p/n \rightarrow \gamma \in (0,1)$ ($n$ is the size of samples), then the empirical spectral distribution of the sample covariance converges to the Mar$\breve{\mbox{c}}$enko-Pastur law $F_\gamma(dx)$
\[
F_\gamma(dx)=\frac{1}{2 \pi x \gamma \sigma^2} \sqrt{(b_\gamma-x)(x-a_\gamma)}dx, \; a_\gamma \leq x \leq b_\gamma
\]
where $a_\gamma=\sigma^2(1-\sqrt{\gamma})^2$, $b_\gamma=\sigma^2(1+\sqrt{\gamma})^2$. This means, with high probability, the $k$-largest eigenvalues (or $k$-smallest eigenvalues) of the sample covariance matrix converge to the spectrum bounds $a_\gamma$ (or $b_\gamma$), rather than the population eigenvalue $\sigma^2$ in the null case. The second reason outlined above implies that the inverse of transition function does not generally work, unless there are only a few distinct spiked eigenvalues, which is usually the case in real applications (in MEG, we expect only a few spiked eigenvalues). A crucial problem, particularly in MEG, is that the magnitude of the unknown noise aggregates the estimation error of the population spikes under the non-Gaussian model, which fails most of existing methods, such as PCA. This calls for a re-examination of the effect caused on the sample extreme eigenvalues by perturbation on the population covariance matrix, the essence of which involves developing new estimates for the population eigenvalues and a justification for its consistency in Gaussian and non-Gaussian spiked model.

The goal of this paper is to bridge the estimation the number of sources and the behavior of the spiked eigenvalues, on the population level. The remainder of the paper is structured as follows: in Section 2, we demonstrate how the problem can be transformed into one of estimating the extreme eigenvalues of the population matrix, instead of directly investigating the limiting behavior of sample spikes. This differentiates our method from other thresholding methods. The relation of the spiked sample eigenvalues and the spiked population eigenvalues is then presented. A new estimator for the population spiked eigenvalues is given, with an algorithm estimating the intrinsic dimensionality under an optimal signal-to-noise ratio transformation. In Section 3, by means of simulated examples and the {\it empty-room} data, we illustrate that our method is able to capture the number of sources under various SNRs, with hundreds of channels or more. We also compare our approach with other methods. Finally, a real MEG data set is tested in Section 4, and a short discussion and concluding remarks are given in Section 5.

\section{Methodology}

In a typical MEG experiment, the magnetic field $\mathbf{B}$ is sampled on a finite number (say $K$) of sensors, with each one measuring one component (radial direction) of the magnetic field, namely $\mathbf{B}_{z}$; %(see the blue arrow in Figure \ref{coil});
if $\mathbf{e}=(0,0,1)$, a unit vector, is used to find $\mathbf{B}_{z}$, the $z$ component of $\mathbf{B}$ can be obtained by $\mathbf{B}_{z}=\mathbf{B}\cdot\mathbf{e}$. Nevertheless, for simplicity, we will ignore the subscript $z$ in $\mathbf{B}_{z}$ from now on. Therefore, the general framework of the MEG model becomes
\begin{equation} \label{Y_t}
\mathbf{Y}(t)=\mathbf{B}(t)+\mathbf{e}(t) =\mathbf{G}\mathbf{Q}(t)+\mathbf{e}(t)
\end{equation}
where $\mathbf{Y}(t)=[Y_1(t), Y_2(t), \cdots, Y_K(t)]^{\top}$ is the $K \times 1$ vector representing the observed magnetic field by $K$ sensors at time $t$. The design matrix $\mathbf{G}=[\mathbf{G}_1, \mathbf{G}_{2}, \cdots, \mathbf{G}_{N}]$ is of size $K \times 3N$ matrix with each submatrix $\mathbf{G}_{i}$ of $K \times 3$ being the corresponding magnetic field observed across sensors generated by a unit dipole at a given location $\mathbf{r}_{i} (1 \leq i \leq N)$. The $\mathbf{Q}(t)=[Q^{\top}_1(t), Q^{\top}_2(t), \cdots, Q^{\top}_N(t)]^{\top}$ is the time course vector representing the strength and moments of all $N$ dipoles, with each submatrix $Q_i(t)$ of $3 \times 1$ being the time course of $i$-th dipole at time $t$. Considering the entire time course of $\mathbf{Q}(t)$, we denote the covariance matrix of $\mathbf{B}(t)$ (or $\mathbf{G}\mathbf{Q}(t)$) as $\mathbf{R}_s$. The vector $\mathbf{e}(t)=[e_1(t), e_2(t), \cdots, e_K(t)]^{\top}$ accounts for the presence of additive noise in the MEG data.

\subsection{Optimal signal-to-noise ratio (SNR) and intrinsic dimensionality (ID)} %The problem of estimating  the value of $N$ in equation (\ref{Y_t}), can be formulated as the following maximization problem. %find the minimum number of parameters required to account for the observed data.
Before discussing the method of determining the number of sources in MEG, introduce a so-called SNR rescaling functional $\mathcal{I}_{\mathbf{R}, \mathbf{R}_n}$, such that for any matrix $\mathbf{X}\in \mathbb{R}^{K \times K}$
\begin{equation}
\mathcal{I}_{\mathbf{R}, \mathbf{R}_n} (\mathbf{X}) = \frac{\|\mathbf{X}^{\top} \mathbf{R} \mathbf{X}\|_2}{\|\mathbf{X}^{\top} \mathbf{R}_n \mathbf{X}\|_2},
\label{functional}
\end{equation}
where $\mathbf{R}$ and $\mathbf{R}_{n}$ are the covariance matrix of the observed data $\mathbf{Y}(t)$ and noise $\mathbf{e}(t)$, respectively; and $\|\mathbf{X}\|_2= \sqrt{\lambda_{max}(\mathbf{X}^{\top}\mathbf{X})}$ is a standard operation norm for a matrix $\mathbf{X}\in \mathbb{R}^{K \times K}$. Such a norm is invariant with respect to any unitary matrix transformation, i.e., for any unitary matrix $U_1,U_2$ the equality $\|U_1\mathbf{X}U_2\|_2=\|\mathbf{X}\|_2$ holds. Therefore, the transformation can be employed for the signal and noise simultaneously. As a ratio between the noise and signal, a high value of $\mathcal{I}_{\mathbf{R}, \mathbf{R}_n}$ means a small effect caused by noise. 

The following theorem finds an optimal transformation for which the SNR rescaling functional $\mathcal{I}_{\mathbf{R}, \mathbf{R}_n}$ attains its maximum. The proof can be found in the Appendix.

\begin{thm}\label{theorem_max}
Suppose that matrix $\mathbf{R}$ is nonsingular. Let $\mathbf{\Phi}_n$ and $\mathbf{\Lambda}_n$ be eigenvectors and eigenvalues of noise covariance $\mathbf{R}_n$. Let $\lambda_{max}>0$ be the maximal eigenvalue of the matrix $\mathbf{R}_{adj}$, where
\begin{equation}
\mathbf{R}_{adj} = \mathbf{W}^{\top}_n \mathbf{R} \mathbf{W}_n, \qquad \mathbf{W}_n = \mathbf{\Phi}_n \mathbf{\Lambda}^{-1/2}_n.
\label{Wn}
\end{equation}
Then
\begin{equation*}
\max_{\mathbf{X}\in \mathbb{R}^{K \times K}} \mathcal{I}_{\mathbf{R}, \mathbf{R}_n} (\mathbf{X}) = \lambda_{max},
\label{max_value}
\end{equation*}
and
\begin{eqnarray}
\mathbf{X}^{opt}_1 = \mathbf{W}_n ,
\label{trans1}
\end{eqnarray}
\begin{eqnarray}
\mathbf{X}^{opt}_2 = \mathbf{\Phi}_n \mathbf{\Lambda}^{-1/2}_n \mathbf{\Phi}_{adj}
\label{trans2}
\end{eqnarray}
are two maximums of functional $\mathcal{I}_{\mathbf{R}, \mathbf{R}_n} (\cdot)$, i.e., $\mathcal{I}_{\mathbf{R}, \mathbf{R}_n} (\mathbf{X}^{opt}_i) = \lambda_{max}$ for $i=1,2$. Here, the columns of matrix $\mathbf{\Phi}_{adj}$ are the eigenvectors of $\mathbf{R}_{adj}$.
\end{thm}

If we assume the signal and noise are uncorrelated as in (\ref{Y_t}), then the covariance matrix $\mathbf{R}$ simply breaks down to
\begin{equation*}
\mathbf{R}=\mathbf{R}_{s}+\mathbf{R}_{n}
\end{equation*}
where $\mathbf{R}_{s}$ is the signal covariance matrix. Suppose we know the noise covariance matrix $\mathbf{R}_n$, then a whitening process can be applied to transform $\mathbf{R}$ and $\mathbf{R}_n$
\begin{equation} \label{associated}
\mathbf{W}^{\top}_n \mathbf{R} \mathbf{W}_n=\mathbf{W}^{\top}_n \mathbf{R}_s \mathbf{W}_n+ \mathbf{I},
\end{equation}
where $\mathbf{W}_n$ is defined in \eqref{Wn}.

Theorem \ref{theorem_max} implies that, with $\mathbf{X}=\mathbf{X}_2^{opt}$ defined in \eqref{trans2}, we can rewrite
\[
\mathbf{X}^{\top} \mathbf{R} \mathbf{X}=\mathbf{\Lambda}_{adj}=\mbox{diag} (\lambda_{1},\lambda_{2},...,\lambda_{M},\lambda_{M+1},...,\lambda_{K})
\]
by noting that $\mathbf{R}_{adj}=\mathbf{W}^{\top}_n \mathbf{R} \mathbf{W}_n$ and the fact that
\[
\mathbf{\Phi}_{adj}^{\top} \mathbf{R}_{adj} \mathbf{\Phi}_{adj}= \mathbf{\Lambda}_{adj}.
\]
By equation \eqref{associated}, it is not difficult to show that $\lambda_k\geq1, k=1, ..., K$. Denote by $M$ the number of nonzero eigenvalues of $\mathbf{R}_s$. Then, we can split $\left\{\lambda_{k}\right\}_{k=1}^{K}$ into two groups: let $\left\{\lambda_{k}\right\}_{k=1}^{M}$ be the first group (we call it the spiked group), where all eigenvalues are strictly larger than the unit; the second group, which is called the bulk group, contains only the unit eigenvalues, i.e., $\left\{\lambda_{k} \equiv 1\right\}_{k=M+1}^{K}$. Obviously, the spiked group can be regarded as the contribution of the associated eigenvalues of both $\mathbf{R}_{s}$ and $\mathbf{R}_{n}$, whereas the ones in the bulk group originate from the eigenvalues of $\mathbf{R}_{n}$ after the whitening process. Therefore, the intrinsic dimensionality (ID) of the data can be determined by counting the number of distinct eigenvalues of $\mathbf{R}_{adj}$ in the spiked group.
 In the next subsection, we confirm that the ID is an invariant, therefore, it can be used to estimate the number of sources in the MEG problem. However, with limited knowledge of $\mathbf{R}_n$, %that is, in general, $\mathbf{R}_n$ is unknown,
one would mainly rely on an estimate of $\mathbf{R}_n$, say $\widehat{\mathbf{R}}_n$. The challenge is that the distribution of the eigenvalues of $\widehat{\mathbf{R}}_{adj}$ are no longer tractable, in the sense that neither does the thresholding apply nor are the sample eigenvalues of $\widehat{\mathbf{R}}_{adj}$ good estimates of their population counterparts. Rather than utilizing  the spiked sample eigenvalues, we propose a new method to estimate the spiked  population eigenvalues of $\mathbf{R}_{adj}$, based on which the inference of the dimensionality is made (see Theorem \ref{estimator} in Section 2.2).

\begin{defi}\label{opti}
$\mathbf{X}^{opt}_i$ ($i=1,2$) defined in \eqref{trans1} and \eqref{trans2}, are called  optimal transformations\footnote{$\mathbf{X}^{opt}_1$ and $\mathbf{X}^{opt}_2$ are essentially equivalent; we will be only using $\mathbf{X}^{opt}_1$.} for the SNR-functional \eqref{functional}. Moreover, $\mathbf{R}_{adj}= \left( \mathbf{X}^{opt}_1 \right)^{\top} \mathbf{R} \mathbf{X}^{opt}_1 \equiv \mathbf{W}^{\top}_n \mathbf{R} \mathbf{W}_n$ is called an optimal associated covariance matrix (with respect to the optimal transformation $\mathbf{X}^{opt}_1$). Accordingly, $\widehat{\mathbf{R}}_{adj}= \widehat{\mathbf{W}}^{\top}_n \widehat{\mathbf{R}}\widehat{\mathbf{W}}_n$, with an estimated $\widehat{\mathbf{R}}$ and $\widehat{\mathbf{W}}_n$ (named as a quasi-optimal transformation), is called a quasi-optimal associated covariance matrix.
\end{defi}

An the end of this subsection, we indicate that it is reasonable to use the quasi-optimal associated covariance matrix $\widehat{\mathbf{R}}_{adj}$ instead of the optimal one $\mathbf{R}_{adj}$, if a good noise estimation has been captured. Theorem \ref{theoremConsistent} (its proof can be found in the appendix) also implies that the transformation used in Theorem \ref{theorem_max} is a stable operator.

\begin{thm}\label{theoremConsistent}
Let $\widehat{\mathbf{R}}_n$ be an $\omega$-estimator of $\mathbf{R}_n$ such that
\begin{equation}\label{ConsistentCondition1}
\|\widehat{\mathbf{\Phi}}_n-\mathbf{\Phi}_n\|\leq \omega, \quad \|\widehat{\mathbf{\Lambda}}^{-1/2}_n-\mathbf{\Lambda}^{-1/2}_n\|\leq \omega,
\end{equation}
where $\{\widehat{\mathbf{\Phi}}_n, \widehat{\mathbf{\Lambda}}_n\}$ and $\{\mathbf{\Phi}_n, \mathbf{\Lambda}_n\}$ are eigenvectors and eigenvalues of $\widehat{\mathbf{R}}_n$ and $\mathbf{R}_n$, respectively. $\|\cdot\|$ denotes any type of norm of a matrix. Moreover, assume that there exists a constant $C$ such that $\|\mathbf{\Lambda}^{-1/2}_n\|\leq C$. Then
\begin{equation*}\label{ConsistentResult}
\|\widehat{\mathbf{R}}_{adj}-\mathbf{R}_{adj}\| = O(\omega).%C \omega+o(\omega). %\mathcal{O} (\omega).
\end{equation*}
\end{thm}
%and $o(\omega)$ is the infinitesimal of higher order

\subsection{Determining the intrinsic dimensionality (ID)}\label{id}
In this section, we present the details to determine the ID from the quasi-optimal associated covariance matrix $\widehat{\mathbf{R}}_{adj}$. % Recall the method of estimating the eigenvalues of a population covariance matrix from the sample covariance matrix in a spiked covariance model and then provide an algorithm for determining the ID. Above all,
We indicate that, from this section, instead of using the original data $\mathbf{Y}(t)$, we will only deal with the data after the quasi-optimal transformation, introduced in the previous section, i.e. the new data structure $\mathbf{Z}(t)=\widehat{\mathbf{W}}^{\top}_n \mathbf{Y}(t)$.

\subsubsection{\textbf{Definition and assumption}}
\begin{defi}
Let $\mathbf{V}_K$ be a population covariance matrix of dimension $K$ and assume that the empirical spectral distribution (ESD) $H_T$ of $\mathbf{V}_K$ tends to a proper probability distribution $H$, called the limiting spectral distribution (LSD), as the dimension $K \to \infty$. An eigenvalue $\lambda$ of $\mathbf{V}_K$ is called a spiked eigenvalue if $\lambda\not\in \Gamma_H$, where $\Gamma_H$ denotes the support of $H$. Otherwise, it is called a bulk eigenvalue.
\end{defi}

{\bf Remark:} To avoid possible confusion when the eigenvalues vary with $K$, this definition can be modified as $d(\lambda(\mathbf{V}_K),\Gamma_H)>\delta_0$ for \emph{spiked eigenvalues} and otherwise for \emph{bulk eigenvalues}, where $\delta_0$ is a pre-chosen positive constant and $d(\cdot)$ is a distance function i.e., $d(\lambda(\mathbf{V}_K),\Gamma_H) = \inf_{x \in \lambda(\mathbf{V}_K), y\in \Gamma_H} |x-y|$.

We generate the spiked covariance model in \ctn{bai2012}. Define $\mathbf{Z}(1), \mathbf{Z}(2), \cdots, \mathbf{Z}(T)$ i.i.d. sample vectors drawn from the $K$-dimensional population with mean vector $\boldsymbol\mu$ and covariance matrix
\begin{equation}
\mathbf{V}_{K} = \mathcal{O}^{\top} \textrm{diag}(\lambda_1 I_{m_1}, \lambda_2 I_{m_2}, \cdots, \lambda_L I_{m_L}, I_{K-M}) \mathcal{O},
\end{equation}
with unit bulk eigenvalues and  $L$ distinct spiked eigenvalues $\{\lambda_l\}^L_{l=1}$, with respective multiplicities $m_1, ..., m_L$, satisfying $\lambda_1>\lambda_2>\cdots>\lambda_L$. Let $M$ be a fixed constant. Denote $\sum^L_{l=1} m_l = M < K$, and assume finite fourth moment $E \|\mathbf{Z}(t)\|^4 < \infty$ ($1 \leq t  \leq T$). Here $I_{m_l}$ is an identity matrix of size $m_l$. Without losing generality, we assume $\boldsymbol\mu=0$. %Therefore, $\mathbf{Z}(t)=\mathbf{X}^T \mathbf{Y}(t)$.

\begin{defi}
The number $L$ will be called the intrinsic dimensionality.
\end{defi}

The intrinsic dimensionality $L$ will be determined by Algorithm 1. Here, throughout, we assume that
\begin{enumerate}
\item[(\textbf{A1})] There exists an orthogonal $K$-dimensional matrix $\mathcal{O}$ such that $\mathbf{Z}(t)= \mathcal{O} \mathcal{Z}(t), t=1, ..., T$ are the sample vectors and sequence $\mathcal{Z}(t)$ can be split as $\mathcal{Z}(t)=(\mathcal{Z}^{\top}_M(t),\mathcal{Z}^{\top}_{K-M}(t))^{\top}$ according to their dimensions $M$ and $K-M$. Moreover, $\mathcal{Z}_M(t)$ and $\mathcal{Z}_{K-M}(t)$ are independent.
\item[(\textbf{A2})] $K$ and $T$ are related so that $K/T\to \gamma \in (0,1)$ as $T\to \infty$.
\item[(\textbf{A3})] $\lambda_M - \lambda_{M+1} > \sqrt{K/T}$, i.e., the gap between the spiked and bulk eigenvalues is larger than a critical value $\sqrt{K/T}$.
\item[(\textbf{A4})] There exists a positive number $\epsilon_0$ such that $\min^{L-1}_{i=1} (\lambda_i - \lambda_{i+1})\geq \epsilon_0$.
\end{enumerate}

Assumption (\textbf{A1}) guarantees the extension of the spiked covariance model in \ctn{bai2012} to the model with a non-block structure (i.e., a dense $\mathbf{V}_K$). This is the well-known ``source condition'' in regularization theory. Assumption (\textbf{A3}) is needed, since the BBP phase transition in the spiked population model exists, which says that only when the population spike is larger than a critical value, will its corresponding sample counterpart have a different asymptotic behavior from the null case (see \ctn{baik2005} for details). In practice, the value of $M$ will be estimated  by Algorithm \ref{learning2}. Assumption (\textbf{A4}) sets the lower bound of the minimum eigen-gap of the spiked population eigenvalues.

\subsubsection{\textbf{Estimating spiked eigenvalues}\\}

To estimate the spiked eigenvalue of the population matrix, define
\begin{equation}
\underline{s}_{T,k} = \frac{\gamma_T - 1}{\hat{\lambda}_{T,k}} + \frac{1}{T} \sum_{j\not\in J_l} \frac{1}{\hat{\lambda}_{T,j}-\hat{\lambda}_{T,k}}, \quad k \in J_l,
\label{s_Tk}
\end{equation}
where $\gamma_T= K/T$, $\hat{\lambda}_{T,k}$ ($k=1, ..., K$) is the $k$-th sample spiked eigenvalues of the quasi-optimal associated covariance matrix $\widehat{\mathbf{R}}_{adj}= \widehat{\mathbf{W}}^{\top}_n \widehat{\mathbf{R}} \widehat{\mathbf{W}}_n$, and $J_l=\{s_l +1, ..., s_l + m_l\}$ is denoted as the index set of the $l$-th population spiked eigenvalue $\lambda_l$, where $s_l= \sum^{l-1}_{i=1} |J_i|$.

As we can see in \eqref{s_Tk}, the value of $\underline{s}_{T,k}$ consists of two parts. The first part is the contribution of the $k$-th eigenvalue of the sample matrix $\widehat{\mathbf{R}}_{adj}$ with the factor $\gamma_T - 1$, and the second part is the contribution of the remaining eigenvalues with the factor of the sample size $T$.

The following theorem provides an estimator of the eigenvalues of a population covariance matrix from the sample covariance matrix, which helps us to estimate the ID.
\begin{thm}[Theorem 3.1, \ctp{bai2012}]\label{estimator}
If $\lambda_l$ is a distant population spike with multiplicity $m_l$, then, under the existence of the $4$-th moment of underlying distributions,
\begin{equation}
-1/\underline{s}_{T,k} \to \lambda_l
\label{theorem2}
\end{equation}
as $T\to\infty$ almost surely, for any $k \in J_l$.
\end{thm}

Note that the limiting property of the estimators $\underline{s}_{T,k}$ relies on the true set $J_l$. However, the exact set of $J_l$ is not accessible in most of the cases. We will change the condition of summation in \eqref{s_Tk} by replacing $\{ j\not\in J_l \}$ with the set $\{ j: |\hat{\lambda}_{T,j} - \hat{\lambda}_{T,k}| >  \varepsilon' \}$. Moreover, theoretically, for any $k \in J_l$, $-1/\underline{s}_{T,k}$ can be considered as an estimator of the spiked population eigenvalue $\lambda_l$. In this paper, using the ``averaging'' technique, instead of choosing one of $\{\underline{s}_{T,k}\}$ we use the following ``averaged'' one
\begin{equation}
\underline{s}^{(l)}_{T} = \frac{\gamma_T - 1}{\hat{\lambda}^{(l)}_{T}} + \frac{1}{T} \cdot \sum_{ \{ j: ~|\hat{\lambda}_{T,j} - \hat{\lambda}^{(l)}_{T}| >  \varepsilon' \}} \frac{1}{\hat{\lambda}_{T,j}-\hat{\lambda}^{(l)}_{T}}.
\label{slT}
\end{equation}
with
\begin{equation}\label{Jl}
J_l= \{ j: |\hat{\lambda}_{T,j} - \hat{\lambda}_{T,k_l}| \leq  \epsilon, ~ k_l< j \leq K\},~ J_0= \emptyset,\\
\end{equation}
\begin{equation}\label{kl-lT}
k_l= 1+ \sum^{l-1}_{i=1} |J_i|, \qquad  \hat{\lambda}^{(l)}_{T} = \frac{1}{|J_l|} \sum_{k\in J_l} \hat{\lambda}_{T,k}\\
\end{equation}
%\begin{equation}\label{lT}
% \hat{\lambda}^{(l)}_{T} = \frac{1}{|J_l|} \sum_{k\in J_l} \hat{\lambda}_{T,k}
%\end{equation}
where $\varepsilon'$ in (\ref{slT}) and $\epsilon$ in (\ref{Jl}) are two thresholds. Note that $\varepsilon'$  can be any arbitrary small number decided by users. This is to remove some eigenvalues closest to $\hat{\lambda}_{T,k}$ in \eqref{s_Tk} (or $\hat{\lambda}^{(l)}_{T}$ in \eqref{kl-lT}). This modification will not affect the consistency of the estimator, as stated in \ctn{bai2012}. We set $\varepsilon' \equiv 1\% \cdot \hat{\lambda}_{T,1}$ throughout this paper. The threshold $\epsilon$, however, is a tuning parameter. It controls the radius of the sample eigenvalues necessary for inclusion in order to estimate the corresponding spiked population eigenvalue.

It is not difficult to show that $-1/\underline{s}^{(l)}_{T} \to \lambda_l$ as $T\to\infty$, and it is a better estimator in comparison with $\underline{s}_{T,k}$ (at least from the numerical point of view- see a demonstration at the end of this subsection).

We first present a high-level description of the algorithm~-- Algorithm \ref{ID1}. The element of noise estimation in Step 2 will be briefly discussed in the next section. Readers may refer to the supplementary materials for more details.  Apart from the spiked population eigenvalues, the bulk population eigenvalues will be estimated simultaneously with tuning parameter $\delta$. The thresholds of $\delta$ and $\epsilon$ in Step 4 are to be determined by a self-exploited procedure. This will be explained through Algorithms \ref{learning1} and \ref{learning2}.  A complete integration of Algorithms \ref{ID1}~-- \ref{learning2} can be found in the supplementary materials.

\subsubsection{\textbf{Determining optimal thresholds $\delta$ and $\epsilon$} \\}

Theoretically, the thresholds $\delta$ and $\epsilon$  can be learned by the following optimization problem
\begin{equation}\label{thresholds}
\{\delta,\epsilon\}= \argmin_{\delta,\epsilon} \| \frac{1}{V} \sum^V_{i=1} \mathbf{Z}_i(\delta,\epsilon) - \bar{\mathbf{Z}} \|^2_F,
\end{equation}
where $\mathbf{Z}=\left[\mathbf{Z}(1),\cdots,\mathbf{Z}(T)\right]$ denotes the transformed MEG data, $\|\cdot\|_F$ means the Frobenius norm of a matrix, and $\mathbf{Z}_i(\delta,\epsilon)$ is a random sample distributed with the mean vector $\bar{\mathbf{Z}}=\frac{1}{T}\sum_{t=1}^{T}\mathbf{Z}(t)$ and covariance matrix $\widehat{\mathbf{V}}_K$ by the estimated spiked population eigenvalues $\tilde{\lambda}_l$'s from Algorithm 1. Alternatively, one may use $\widehat{\mathbf{V}}'_K=\widehat{\mathcal{O}}^{\top}\widehat{\mathbf{V}}_K \widehat{\mathcal{O}}$, where $\widehat{\mathcal{O}}$ comes from the spectral decomposition of the sample covariance matrix $\widehat{\mathbf{R}}=\widehat{\mathcal{O}}^{\top} \mathbf{\Lambda} \widehat{\mathcal{O}}$. It is enough to use $\widehat{\mathbf{V}}_K$ for optimizing $\epsilon$ in \eqref{thresholds}, since $\widehat{\mathbf{V}}_K$ contains nearly all information about the eigenvalues. Note that the estimated bulk population eigenvalues can be set as the average of the sample bulk eigenvalues, i.e.,

\begin{equation}\label{bulk}
\tilde{\lambda}_0 = \frac{1}{|\Gamma|} \sum_{\hat{\lambda}_{T,k}\in \Gamma} \hat{\lambda}_{T,k},
\end{equation}
where set $\Gamma :=\{\hat{\lambda}_{T,k} :~ \hat{\lambda}_{T,k}\leq \delta, k=1, ..., K\}$ contains all bulk sample eigenvalues of $\widehat{\mathbf{R}}_{adj}$ and $|\Gamma|$ denotes as the cardinality of set $\Gamma$.

%\begin{equation}\label{bulk}
%\mathbf{E} \left( \{\hat{\lambda}_{k}: \hat{\lambda}_{k}\leq \delta\} \right).
%\end{equation}
%where the $\mathbf{E}$ is taken over all the $K-L$ ($1 \leq k \leq K$) eigenvalues of  %$\widehat{\mathbf{R}}_{adj}$ excluding the spiked eigenvalues.

A detailed implementation of learning process \eqref{thresholds} is presented in Algorithms \ref{learning1} and \ref{learning2}, where Algorithm \ref{learning1} presents an iterative method of selecting the optimal threshold $\epsilon$, and Algorithm \ref{learning2} shows an efficient approach for choosing an appropriate threshold $\delta$.

\vspace{0.5cm}
\scalebox{0.78}{
\begin{algorithm}[H]
\caption{A high-level algorithm determining the ID}
\label{ID1}
\begin{enumerate}
\item Obtain the quasi-optimal associated covariance matrix $\widehat{\mathbf{R}}_{adj}$ from data covariance $\widehat{\mathbf{R}}$. If the data has a negligible noise, set $\widehat{\mathbf{R}}_{adj}=\widehat{\mathbf{R}}$ and go to Step 3; otherwise, go to the next step.

\item Given any estimated $\widehat{\mathbf{R}}_n$, calculate the quasi-optimal transformation $\widehat{\mathbf{W}}_n$ and obtain the corresponding quasi-optimal associated covariance matrix $\widehat{\mathbf{R}}_{adj}$.

\item Calculate all eigenvalues of $\widehat{\mathbf{R}}_{adj}$: $\{\hat{\lambda}_{T,k}\}^K_{k=1}$ and sort them in descending order.
\item Choose appropriate thresholds $\delta$ and $\epsilon$.
\item Estimate the spiked  population   eigenvalues $\{\tilde{\lambda}_i\}^{L}_{i=1}$ using \eqref{theorem2}~-- \eqref{kl-lT} in Theorem \ref{estimator} from the  sample spiked eigenvalues $\{\hat{\lambda}_{T,k} : \hat{\lambda}_{T,k}> \delta, k=1, ..., K\}$.
\end{enumerate}
{\bf Output}: $\{\tilde{\lambda}_i\}^{L}_{i=1}$ are the estimated population spiked eigenvalues and the ID equals $L$.
\end{algorithm}
}
\vspace{0.5cm}

%{\bf Remark}

\vspace{0.5cm}
\scalebox{0.78}{
\begin{algorithm}[H]
\caption{Determining the optimal threshold $\epsilon$}
\label{learning1}
{\bf Input}: Set the iteration index $j=1$ and provide an appropriate threshold $\delta$ and an initial threshold $\epsilon=\epsilon_0$.
\begin{enumerate}
\item Estimated the spiked population eigenvalues $\{\tilde{\lambda}_i\}^{L}_{i=1}$  from the spiked sample eigenvalues $\{\hat{\lambda}_{T,k}: \hat{\lambda}_{T,k}> \delta, k=1, ..., K\}$ with threshold $\epsilon$.
\item Generate $V$ random samples $\{\mathbf{Z}_i\}^V_{i=1}$ distributed with the mean vector $\bar{\mathbf{Z}}$ and covariance matrix $\widehat{\mathbf{V}}_K$ by the estimated spiked population eigenvalues $\{\tilde{\lambda}_i\}^{L}_{i=1}$.
\item Compute the relative discrepancy $\Delta_j=\|\frac{1}{V} \sum^V_{i=1} \mathbf{Z}_i(\delta,\epsilon)-\bar{\mathbf{Z}}\|_F / \|\bar{\mathbf{Z}}\|_F$.
\item If $\epsilon>\hat{\lambda}_{T,1} - \hat{\lambda}_{T,K}$, output the result; otherwise, update $\epsilon=\epsilon+\epsilon_0$, $j=j+1$ and go back to Step 1.
\end{enumerate}
{\bf Output}: The optimal threshold $\tilde{\epsilon}$ is the one with the minimum relative discrepancy $\Delta_j$.
\end{algorithm}
}
\vspace{0.5cm}

{{\footnotesize {\bf Remark:}   (a) In Algorithm 2, $\epsilon_0$ can be any appropriate value. Obviously, the smaller $\epsilon_0$, the better estimated $\epsilon$. However, small $\epsilon_0$ leads to more iterations in the algorithm. If there are no other good choices, we suggest using $\epsilon_0=\hat{\lambda}_{T,K}$, the smallest sample eigenvalue. (b) By \eqref{Jl}, $\epsilon$ is the radius of set $J_l$, which determines the estimation of population spiked eigenvalue $\hat{\lambda}_{T,l}$. The stopping criteria in Step 4 describes the maximal possible value of the radius of $J_l$. $\epsilon=\hat{\lambda}_{T,1} - \hat{\lambda}_{T,K}$ means the extreme case, when there is only one spiked eigenvalue and one bulk eigenvalue. All sample eigenvalues except the smallest one are used to estimate the population spiked eigenvalue. If the data covariance matrix has several distinguishable spiked eigenvalues, we may use the alternative stopping criteria ``$\epsilon >p \cdot \hat{\lambda}_{T,1}$'', where $p$ should be chosen from case to case. Usually, $p$ should be greater than $30\%$. Obviously, the smaller $p$ is, the fewer iterations it requires. Note that we have used ``$\epsilon >40\% \cdot \hat{\lambda}_{T,1}$''. The two criteria provide the same result in our simulation and real data application.}

\scalebox{0.78}{
\begin{algorithm}[H]
\caption{Choosing an appropriate threshold $\delta$}
\label{learning2}
\begin{enumerate}
\item Set index $i=K$ and $\delta=\hat{\lambda}_{T,K}$.
\item If condition $\hat{\lambda}_{T,i-1}-\hat{\lambda}_{T,i}<\epsilon_0/2$ holds, go to the next step; otherwise, output the result.
\item Set $i=i-1$ and $\delta=\hat{\lambda}_{T,i}$. If $i<2$, output the error information ``The spiked eigenvalues model cannot be employed''; otherwise, go back to Step 2.
\end{enumerate}
{\bf Output}: $\tilde{\delta}=\delta$ is an appropriate threshold.\\
\end{algorithm}
}

{\footnotesize {\bf Remark:} In Algorithm 3, the initial value of $\delta$ is suggested to be $\hat{\lambda}_{T,K}$. The algorithm separates the bulk and spiked eigenvalues and therefore it is an approach to estimate the number of spiked eigenvalues $M$.}

\subsubsection{\textbf{A demonstration of Algorithms 1, 2 and 3}\\}

We first generate a random (Gaussian) sample matrix ($300\times 6000$, i.e., $K=300$ and $T=6000$) with a mean zero and a covariance matrix
\begin{equation}
\mathbf{V}_{300} = \textrm{diag}(\lambda_1 I_{20}, \lambda_2  I_{10}, \lambda_3  I_{40}, \lambda_4  I_{30}, I_{200}),
\end{equation}
where the $\lambda_1=20, \lambda_2=17, \lambda_3=10, \lambda_4=7$ are the four true spiked population eigenvalues. Then, the eigenvalues of the sample covariance matrix $\widehat{\mathbf{R}}$ (in ``$\circ$'') are displayed in Figure \ref{Fig:ExplainSpiked6}. In the same figure, the exact spiked population eigenvalues (in ``$\square$'') and the estimated spiked population eigenvalues (in ``$\times$'', ``$+$'' and ``$\ast$'') by different sampling distributions in Step 2 of Algorithm 2 are also highlighted. The learning process of selecting the optimal threshold $\epsilon$ with the initial guess $\epsilon_0=10\% \cdot \hat{\lambda}_{T,1}$ is displayed in Figure \ref{Fig:ExplainSpikedAll} (the Gaussian distribution is used in Step 2 of Algorithm 2). The result of the learning process is the histogram (c) with the smallest relative discrepancy, i.e., the optimal threshold $\tilde{\epsilon}=3\cdot \epsilon_0= 30\% \cdot \hat{\lambda}_{T,1}$. The result also suggests that the Algorithm 1 recovers the true population spiked eigenvalues well in terms of number and magnitude; specifically, the optimal estimated spiked population eigenvalues are $(20.42,16.85,10.53,7.9)$, which implies that the number of spiked eigenvalues equals four.

\begin{figure}
\centerline{\includegraphics[width=0.5\textwidth]{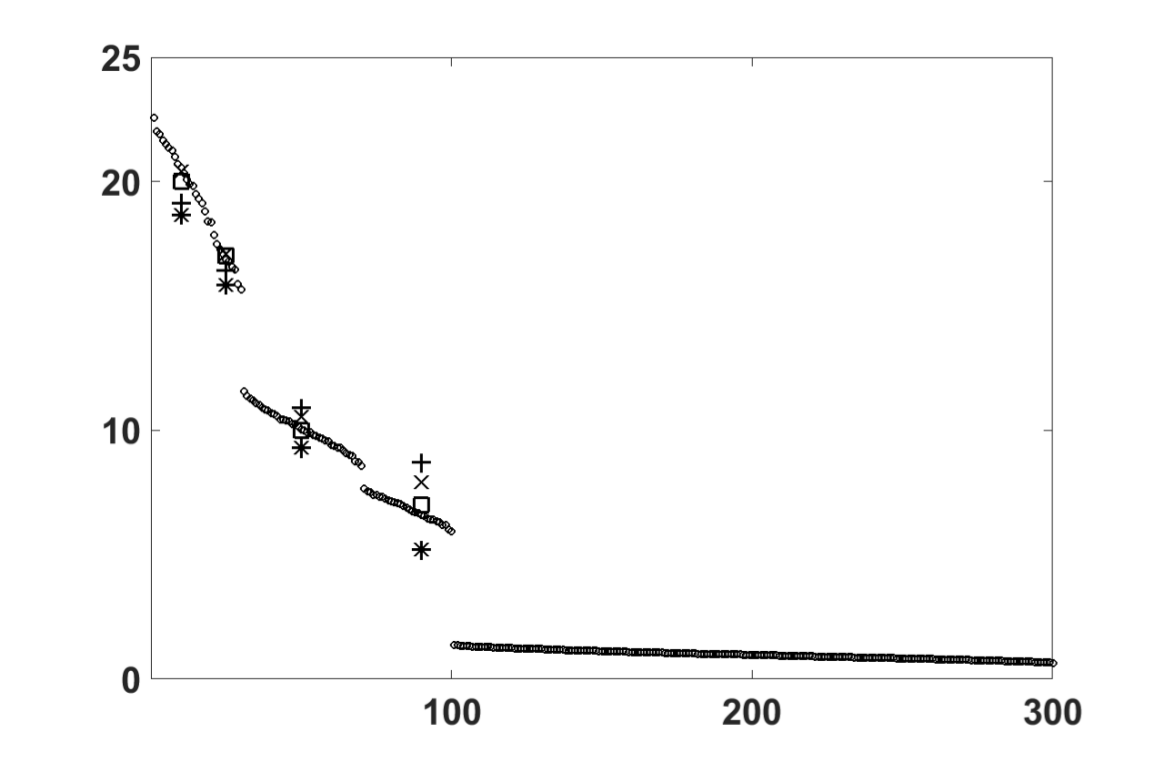}}
\caption[A demonstration of Algorithms 1~-- 3 with different distributions]{A demonstration of Algorithms 1~-- 3 with different distributions in Step 2 of Algorithm 2. ``$\circ$'' represents the sample eigenvalues and ``$\square$'' represents the true spiked population eigenvalues. ``$\times$'', ``$+$'' and ``$\ast$''-- the estimated spiked population eigenvalues with normal distribution, uniform distribution and $t$-distribution, respectively.}
\label{Fig:ExplainSpiked6}
\end{figure}

%\newpage

\begin{figure}[H]
\begin{center}
\subfigure[]{
\includegraphics[width=2.5in]{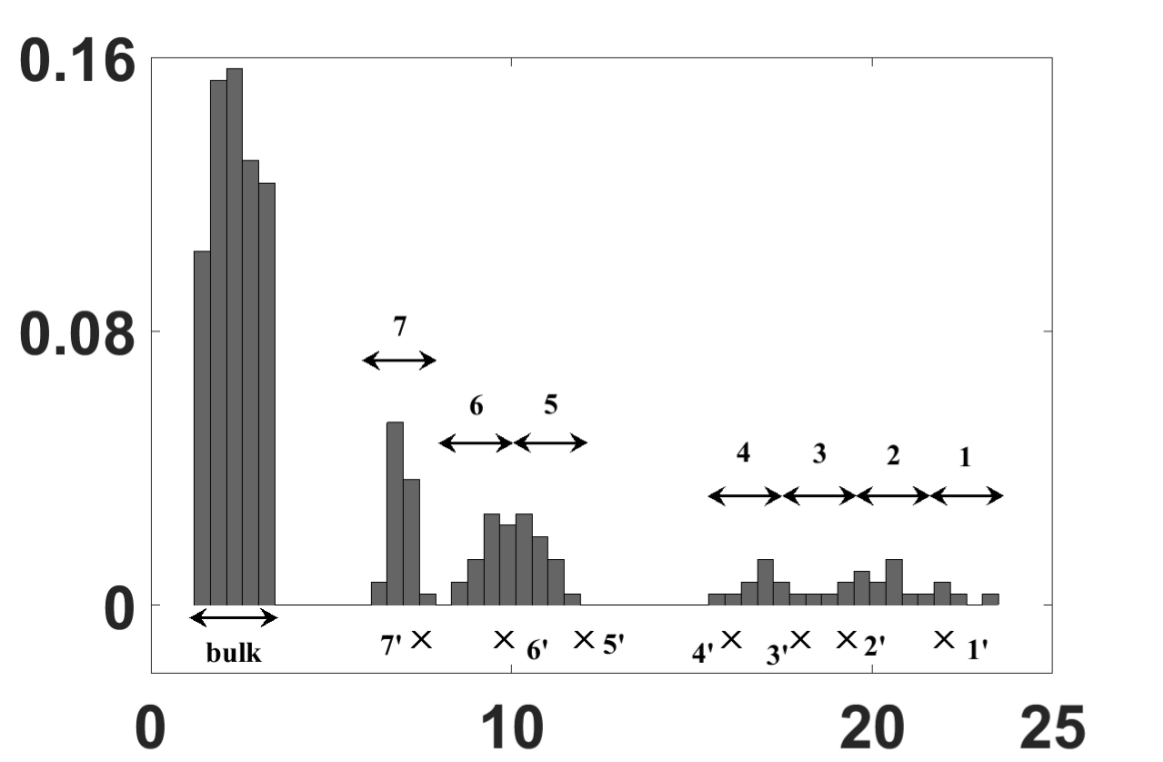}}
%\hspace{.5in}
\subfigure[]{
\includegraphics[width=2.5in]{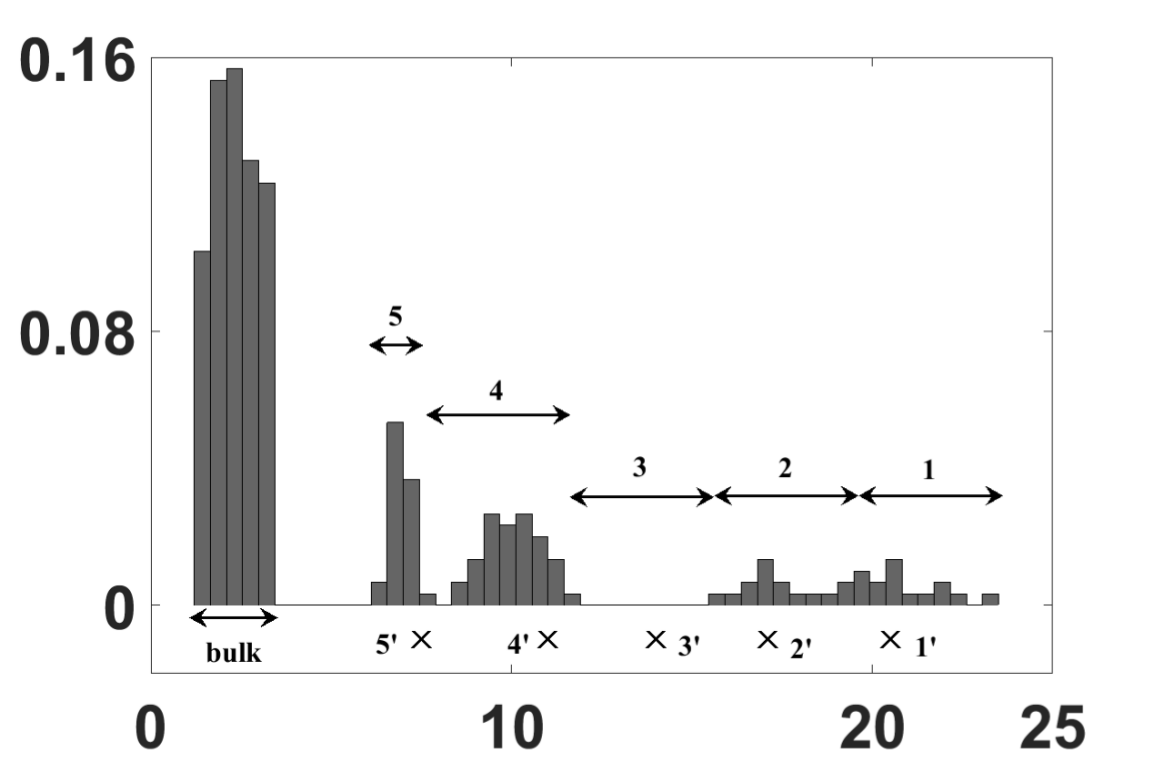}}
\subfigure[]{
\includegraphics[width=2.5in]{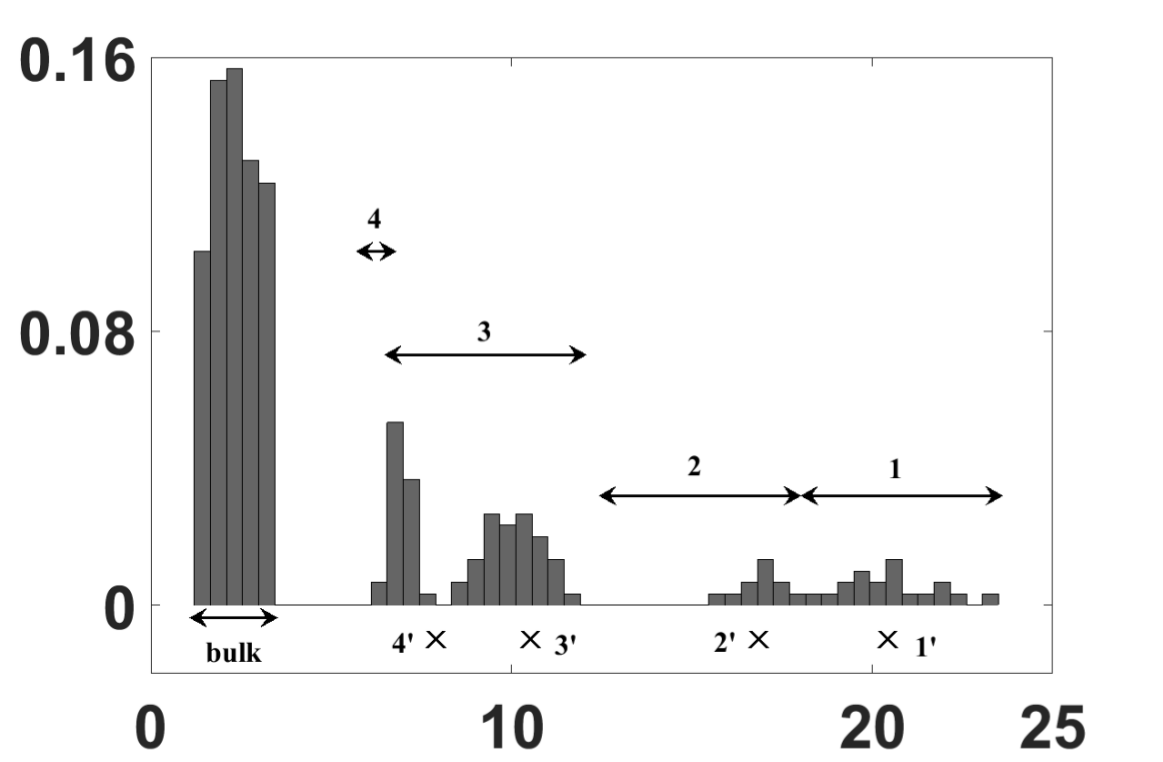}}
\subfigure[]{
\includegraphics[width=2.5in]{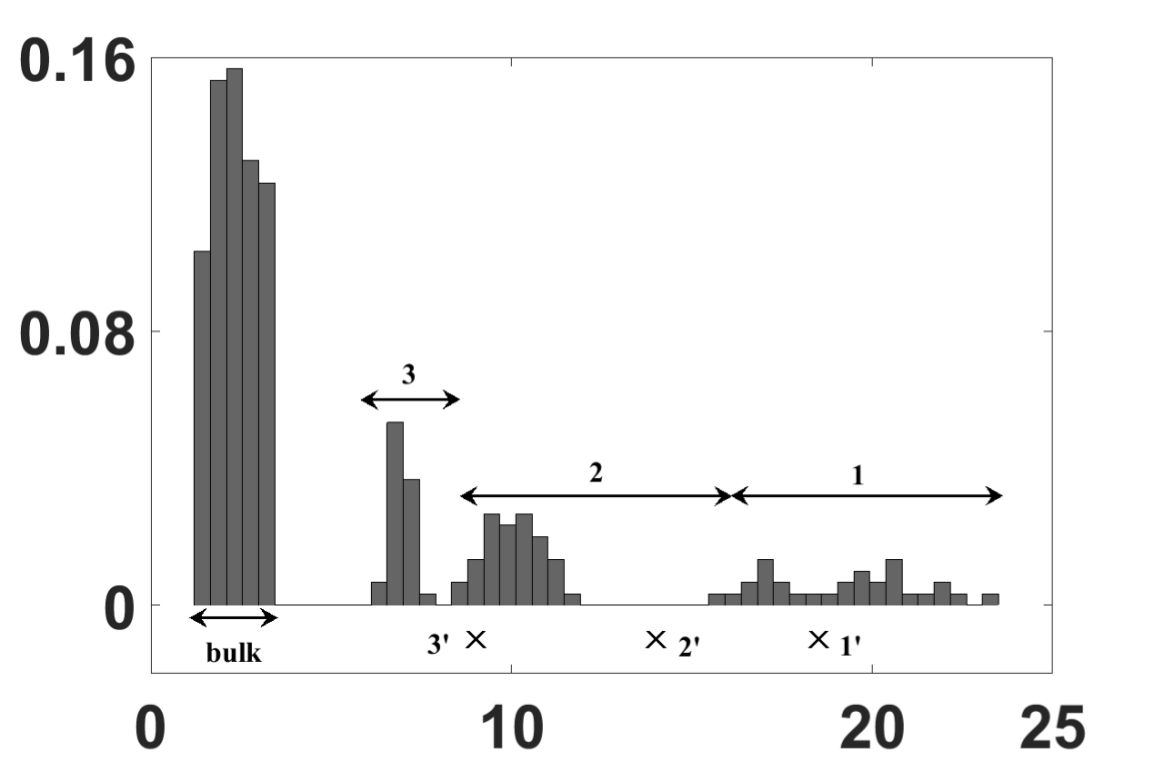}}
\end{center}
\caption[The learning process of selecting the optimal threshold $\epsilon$]{The learning process of selecting the optimal threshold $\epsilon$ with the initial guess $\epsilon_0=10\% \cdot \hat{\lambda}_{T,1}$ and stopping criteria $\epsilon >40\% \cdot \hat{\lambda}_{T,1}$. (a) Results after the first iteration. $\Delta_1=1.67$. (b) Results after the second iteration. $\Delta_2=1.42$. (c) Results after the third iteration. $\Delta_3=1.13$. (d) Results after the forth iteration. $\Delta_4=1.26$. In each histogram, all the double arrows remain the same length (except the last one, since all remaining eigenvalues belong to a class with a smaller threshold), which is the value of the threshold $\epsilon$ in the current iteration. Moreover, the number above each double arrow labels the group of the corresponding spiked eigenvalue, which contains the sample eigenvalues within a distance of the double arrow length. ``$\times$'' with labels $1'$, $2'$, $\cdots$ representing the estimated population spiked eigenvalues, corresponding groups with labels $1$, $2$, $\cdots$.}
\label{Fig:ExplainSpikedAll}
\end{figure}

However, there could be certain variation in the estimated population spiked eigenvalues with different sampling distributions used in tuning $\epsilon$ in Algorithm 2. We observe that in Table \ref{demonstration}, with different initial guesses $\epsilon_0$ in the first step and distributions in Step 2 of Algorithm 2, we have obtained only slightly different spiked population eigenvalues. However, the number of estimated spiked population eigenvalues is very stable with respect to these parameters. The supplementary materials contain more numerical experiments, including several replicates of the same setup, and a case of eight dipoles. In all experiments, the number of estimated spiked population eigenvalues are strictly equal to the truth, although the actual estimated spiked eigenvalues may differ due to the randomness. Therefore, it is reasonable to use it for estimating the sources. Hence, we propose that

\begin{enumerate}
\item[(\textbf{A5})] The number of sources in MEG equals the intrinsic dimensionality of the data.
\end{enumerate}

\begin{table}[!htb]
\caption[A Demonstration of Algorithm 2 with different initial guesses and sampling distributions]{A demonstration of Algorithm 2 with different initial  guesses $\epsilon_0$ and sampling distributions used in Step 2 of Algorithm 2. Each vector with four elements in the table denotes the estimated spiked population eigenvalues (SPEs) with the corresponding initial guess $\epsilon_0$ and the distribution.}
\centering
\begin{tabular}{c c c c} % centered columns (4 columns)
\hline\hline %inserts double horizontal lines
Estimated  SPEs & Gaussian & Uniform & $t$-distribution \\ [0.5ex] % inserts table
%heading
\hline % inserts single horizontal line
$\epsilon_0=5\%\cdot\hat{\lambda}_{T,1}$  & $(19.81,17.09,9.89,6.78)$&$(20.60,16.74,10.38,7.50)$&$(19.15,16.28,9.64,6.40)$  \\ [0.5ex] % inserts table
%$r$ (dm) & 0   & 0.1 & -0.1 & 0.4  \\ [0.5ex] % inserts table
$\epsilon_0=10\%\cdot\hat{\lambda}_{T,1}$  & $(20.42,16.85,10.53,7.9)$&$(19.13,16.43,10.91,8.71)$&$(18.65,15.85,9.32,5.22)$ \\
$\epsilon_0=15\%\cdot\hat{\lambda}_{T,1}$  & $(20.43,17.64,11.03,8.93)$&$(19.42,18.32,9.66,5.95)$&$(17.84,15.14,8.90,6.13)$ \\
 % [1ex] adds vertical space
\hline %inserts single line
\end{tabular}
\label{demonstration}
\end{table}

%{\color{red}
\subsection{Estimation of Noise Covariance}
%The goal of this transform
Consider the transform $\mathbf{Y}(t) \mapsto \frac{1}{\sqrt{T}}  \Omega_n \mathbf{Y}(t)$,
%$\mathbf{Y}(t) \sim N(E(\mathbf{Y}(t)), \mathbf{R})$%
%where $\mathbf{R}=\mathbf{R}_s+\mathbf{R}_n$.
used to suppress the effect of noise from the model \eqref{Y_t}.  If $\Omega_n=I$, the above transformation is the so-called Brute-force transformation (BT). Denote the concentration matrix $\Omega_n=\mathbf{R}^{-1}_n$. If we have a reasonably good estimator $\hat{\Omega}_n$ (or $\widehat{\mathbf{R}}_n$), then $\mathbf{Y}(t) \mapsto  \frac{1}{\sqrt{T}}\hat{\Omega}_n^{1/2} \mathbf{Y}(t)$; this is the whitening transformation (WT) that we have used in \eqref{associated}. %The alternative transformation $\mathbf{Y}(t) \mapsto \frac{1}{\sqrt{n}} \hat{\Omega}_n \mathbf{Y}(t)$ is connected to the term of {\it Innovation} transformation (IT) in the literature (see e.g., \ctn{HallJin2010}).
This being said, challenges remain on how to obtain a good $\hat{\Omega}_n$. Our heuristic is that if we can estimate $\mathbf{R}_n$ reasonably well, particularly the diagonal of the $\mathbf{R}_n$, the resultant de-noised covariance $\widehat{\mathbf{R}}_{adj}$ can be used as a good candidate for finding spiked eigenvalues. In this paper, we mainly consider $\hat{\Omega}_n=\widehat{\mathbf{R}}_n^{-1}$, given an estimate $\widehat{\mathbf{R}}_n$. Note that the estimation might vary case by case; only WT is considered. We are now in position to discuss three algorithms of noise estimation accommodating the three kinds of correlations that possibly exist in the data: inner-sensor, inter-sensor and the combined one.

The first noise estimation algorithm to use is Fourier transform. The idea is to estimate the noise variance on its frequency domain while the MEG data is sampled in the time domain. The second algorithm is based on residual analysis  \cite{Roger_1996}. We estimate the inverse of noise variance instead of the original one. The last algorithm aims to estimate the noise covariance by thresholding methods \cite{Bickel2008}, which has been recently employed on MEG data by \ctn{zhang2015}. Our updated approach results in a non-diagonal noise covariance matrix. A detailed description of the above three noise estimation algorithms can be found in supplementary materials.

\subsection{AIC, MDL and Malinowski's Method}\label{aicmdl}
The information-theoretic criteria AIC \ctp{Wax_1985} and MDL \ctp{Schwarz1978} will be used for comparison. Determining the number of signals is equivalent to finding the number of free parameters in the model \eqref{Y_t}. For a fair comparison, we will evaluate AIC and MDL based on the eigenvalues of the $\widehat{\mathbf{R}}_{adj}$,
%\begin{eqnarray*}
\begin{align*}
 \mbox{AIC}(N)&= -2\log \left(\frac{\prod_{j=N+1}^{K}\hat{\lambda}_{T,j}^{\frac{1}{K-N}}}{\frac{1}{K-N}\prod_{j=N+1}^{L}\hat{\lambda}_{T,j}}\right)^{(K-N)T}+2N(2K-N) \\
     \mbox{MDL}(N)&= -\log \left(\frac{\prod_{j=N+1}^{L}\hat{\lambda}_{T,j}^{\frac{1}{K-N}}}{\frac{1}{K-N}\prod_{j=N+1}^{K}\hat{\lambda}_{T,j}}\right)^{(K-N)T}+\frac{1}{2}N(2K-N)\log T,
\end{align*}
%\end{eqnarray*}
%\begin{align*}
%\mbox{AIC}(N)= -2\log \left(\frac{\prod_{j=N+1}^{K}\hat{\lambda}_{T,j}^{\frac{1}{K-N}}}{\frac{1}{K-N}\prod_{j=N+1}^{L}\hat{\lambda}_{T,j}}\right)^{(K-N)T}+2N(2K-N)
%\end{align*}
%and
%\begin{align*}
%\mbox{MDL}(N)= -\log \left(\frac{\prod_{j=N+1}^{L}\hat{\lambda}_{T,j}^{\frac{1}{K-N}}}{\frac{1}{K-N}\prod_{j=N+1}^{K}\hat{\lambda}_{T,j}}\right)^{(K-N)T}+\frac{1}{2}N(2K-N)\log T
%\end{align*}
where $N$ is the number of free parameters. In our case, $N$ refers to the number of signal sources. If the noise is independent and identically distributed, the problem of finding the number of signal sources can be achieved by minimizing,
\begin{align*}
%\mbox{Number of sources}=\mbox{argmin}\hspace{0.05 in} \left\{\mbox{min}_{p}\hspace{0.05 in} \mbox{AIC}(p)\right\}
\mbox{Number of sources}&=\argmin_{N}\hspace{0.05 in}\mbox{AIC}(N)\\
%\mbox{Number of sources}=\mbox{argmin}\hspace{0.05 in} \left\{\mbox{min}_{p}\hspace{0.05 in} \mbox{MDL}(p)\right\}
\mbox{Number of sources}&=\argmin_{N}\hspace{0.05 in}\mbox{MDL}(N).
\end{align*}

Malinowski's method \ctp{Malinowski_1977_2}, a popular factor analysis method, is also used here for comparison, where an empirical indicator function (EIF) \ctp{Malinowski_1977_1} is introduced as a criterion
\begin{align*}
\mbox{EIF}(N)=
\frac{\left(\sum_{j=N+1}^{K}\hat{\lambda}_{T,j}\right)^{1/2}}{T^{1/2}\left(K-N\right)^{3/2}}
\end{align*}
and the number of sources is estimated by
\begin{align*}
%\mbox{Number of sources}=\mbox{argmin}\hspace{0.05 in} \left\{\mbox{min}_{p}\hspace{0.05 in} \mbox{EIF}(p)\right\}
\mbox{Number of sources}=\argmin_{N}\hspace{0.05 in}\mbox{EIF}(N).
\end{align*}

 Each of the AIC, MDL and EIF methods tend to overestimate the number of signal sources, since they rely on the independence and normality assumption. In Section \ref{2simulation}, their performances are compared with our method.

%\newpage
\section{Simulation Study}
\label{2simulation}
\subsection{Computer Simulation}

Before running our algorithms on a real data set, we tested a simplified case. In this example, we created a channel-level MEG data using a dipole configuration with four dipoles at specified locations in the head.  The location and moments parameters of these simulated dipoles are summarized in Table \ref{2tb1}. We simulated 128 electrodes (magnetometers) by randomly placing them on the upper part of the unit sphere with a radius of 100 mm. The head was modeled by a concentric 3-sphere volume conductor. %The radius is the same as for the electrodes.
The radii of 3 spheres in the conductor is, respectively, $88, 92$, and $100$mm, with its corresponding conductivity $1, 1/80, 1$. The geometrical information and the simulated signal are visualized in Figure \ref{sensors}. Note that the associated parameters for each dipole, such as the locations, did not vary during the simulation. In other words, each dipole contributed a different but constant signal at the same sensor. However, to work with time-varying dipoles, we applied a different frequency to the magnitudes of each dipole so that we could create a distinct time course for each dipole. The time course of each dipole with unique frequency was modeled by either a sine or cosine function, as follows,
\begin{eqnarray*}
\sin (\frac{2 \pi}{1/10} t), \quad \cos (\frac{2 \pi}{1/15} t), \quad \sin (\frac{2 \pi}{1/20} t-\frac{\pi}{4}), \quad \cos (\frac{2 \pi}{1/30} t-\frac{\pi}{4}),
\end{eqnarray*}
where $1 \leq t \leq T$. The pure magnetic signal produced by each dipole at each sensor was calculated using the Biot-Savart equation. The total length ($T$) of each trial is $1000$ timesteps. The magnetometer data were obtained by adding up the contributions from each dipole and the simulated noise across all sensors. To work with different noise levels, we control
\[
\mbox{SNR}=\sigma^2_{{\footnotesize\mbox{signal}}}/{\sigma^2_{{\footnotesize \mbox{noise}}}} %\mbox{SNR}=\frac{\sigma^2_{{\footnotesize\mbox{signal}}}}{\sigma^2_{{\footnotesize \mbox{noise}}}}
\]
where $\sigma^2_{{\footnotesize \mbox{signal}}}$ and $\sigma^2_{{\footnotesize \mbox{noise}}}$ are the variance of the signal and the noise within each trial, respectively. Here, we essentially have $\mathbf{X}=I$ and $\mbox{SNR}=\mathcal{I}_{R,R_n}(I)$.

\begin{figure}[h]
\begin{center}
\subfigure[]{
\includegraphics[width=2.3in]{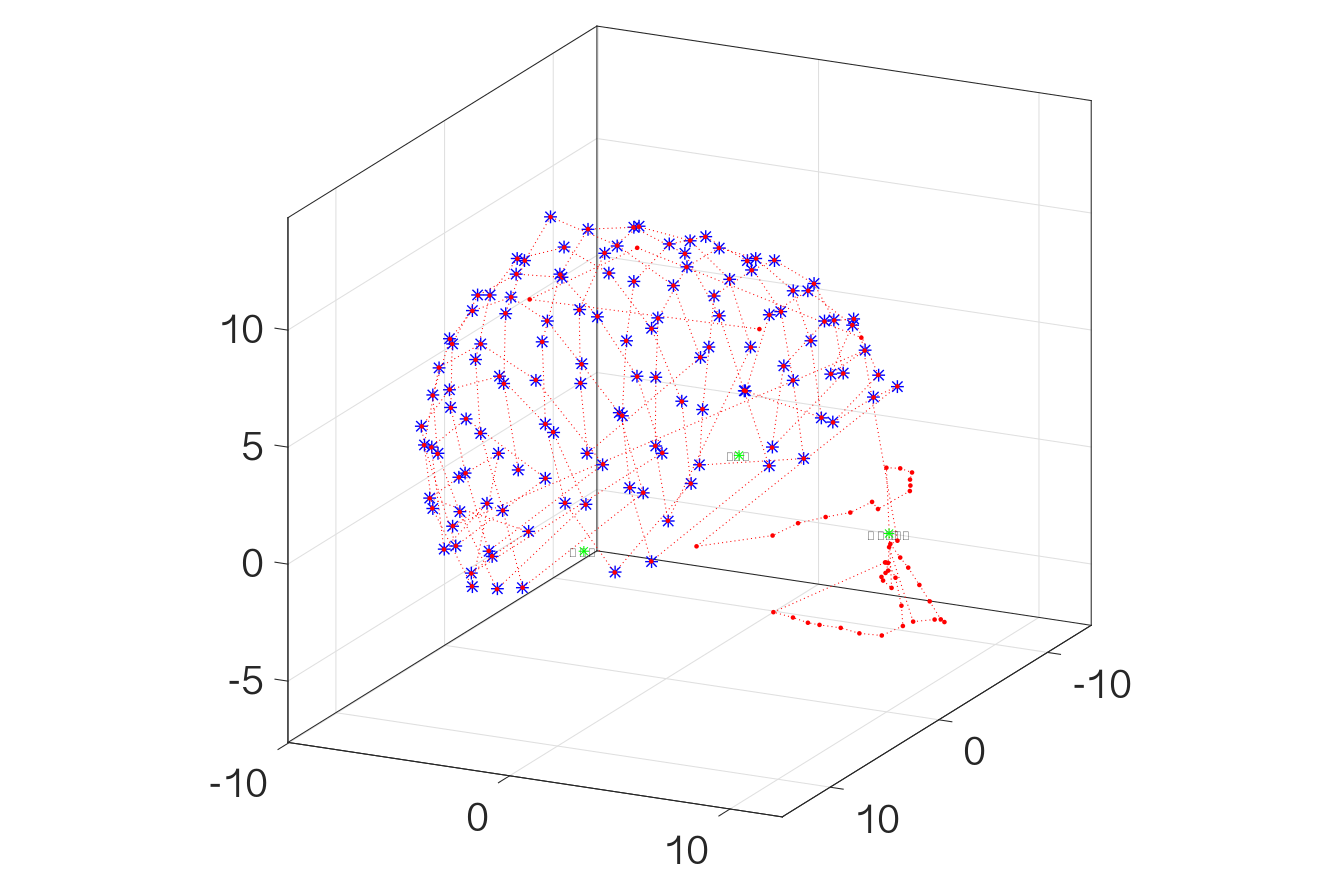}}
\hspace{.1in}
\subfigure[]{
\includegraphics[width=2.3in]{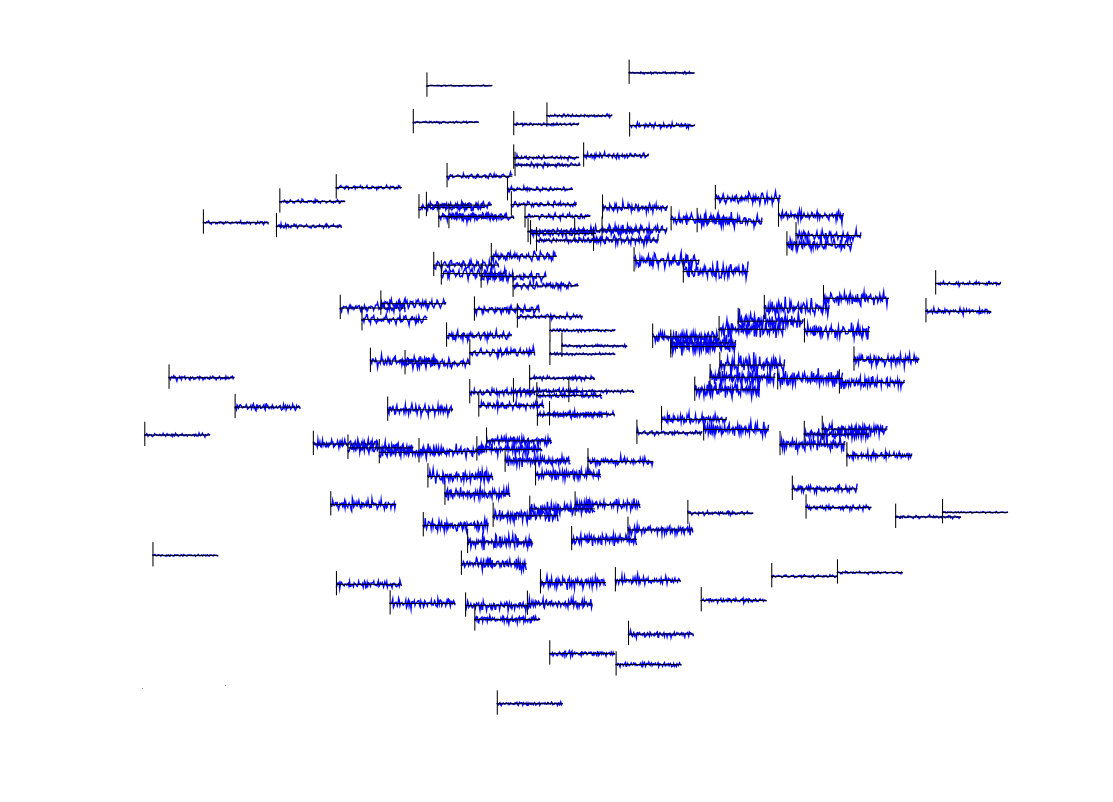}}
\end{center}
%\footnotesize{Figure 6: PVM Structure (Master and Workers' jobs)}
\caption[The sensor map and the simulated signal]{The sensor map and the simulated signal. (a) The simulated MEG sensors on the skull. (b) The simulated magnetic signal generated by the four dipoles at various sensors (vertical view).}
\label{sensors}
\end{figure}

We generated five trials of the data, as follows. In each trial, we calculated the variance of pure signal $\sigma^2_{{\footnotesize\mbox{signal}}, k}$ at each sensor $k$ ($1 \leq k \leq K$). Then, for a given SNR ($1, .1, .01, .001, .0001$), we set the corresponding variance of noise by $\sigma^2_{{\footnotesize \mbox{noise}}, k}= \sigma^2_{{\footnotesize\mbox{signal}}, k}/\mbox{SNR}$. Furthermore, by adding Gaussian noise components with the calculated variance to the simulated data, we obtained the noised signal with an expected SNR in each trail. %This procedure was repeated for five times.
Finally, we averaged the corresponding noised signals over the five trials. Our method, based on the spiked population eigenvalue (SPE), was tested against PCA and other methods such as AIC, MDL and EIF on this averaged data.

\begin{table}[htbp]
\caption[Illustration of dipole simulation]{Illustration of dipole simulation. The location of each dipole (a total of four) is expressed in terms of spherical coordinates ($r$,$\theta$,$\phi$), where $r$ is radial distance, $\theta$ is inclination and $\phi$ is azimuth. $m_{1}$ and $m_{2}$ are the dipole moment parameters. $s$ is the strength parameter of a dipole.}
\centering % used for centering table
\begin{tabular}{c c c c c} % centered columns (4 columns)
\hline\hline %inserts double horizontal lines
Dipole index & 1 & 2 & 3 & 4\\ [0.5ex] % inserts table
%heading
\hline % inserts single horizontal line
$r$ (mm)  & 0   & 10 & -10 & 40  \\ [0.5ex] % inserts table
%$r$ (dm) & 0   & 0.1 & -0.1 & 0.4  \\ [0.5ex] % inserts table
$\phi$   & 0.5 & 0.1 & -0.5 & -0.3 \\
$\theta$ & 3 & 0.1 & -0.3 & 0.3 \\
$m_{1}$ & 1 & 0   & 0.3 & 1  \\
$m_{2}$ & 0 & 0.5 & 0.4 & 0.7 \\
$s$ (mA)& 0 & 0.5 & 0.2 & 0  \\
 % [1ex] adds vertical space
\hline %inserts single line
\end{tabular}
\label{2tb1}
\end{table}

The performance of each method is summarized in Table \ref{2tb3}, in terms of the estimated number of dipoles. We can see that SPE successfully recovers the correct number of dipoles by estimating the number of spiked population eigenvalues, and it outperforms all other methods regardless of the SNR levels. PCA seems to pick up a number of dipoles which is more or less accurate when the SNR is large, but it tends to detect more dipoles for small SNRs. This phenomenon was expected, as the sample eigenvalues of the covariance matrix become increasingly unreliable as a measure of the number of dipoles when the SNR decreases. To illustrate this, we have zoomed in on the distribution of the sample eigenvalues (in ``$\circ$'') and the spiked population eigenvalues (in ``$\ast$'') in Figure \ref{spk}, for SNR=$1, .1, .01, .001$, where the sample eigenvalues of the covariance matrix $\widehat{\mathbf{R}}$ and the quasi-optimal transformed covariance matrix $\widehat{\mathbf{R}}_{adj}$ are displayed, respectively.
From Figure \ref{spk}, we also see that PCA utilizes the sample eigenvalues that become closer and closer in magnitude, either from $\widehat{\mathbf{R}}$ or $\widehat{\mathbf{R}}_{adj}$, and thus it tends to overestimate the number of dipoles. The spiked population eigenvalues of $\widehat{\mathbf{R}}_{adj}$ estimated by SPE are superimposed, accordingly. It is suggested that, as the SNR decreases, the role of spiked population eigenvalues becomes dramatically more significant. In particular, when the SNR=.001, it is impossible to separate the sample eigenvalues, while the SPE still finds the right number of dipoles. Both the AIC and MDL largely overestimate the dipoles across all SNRs, while the number of dipoles estimated by EIF shows a reasonable range that covers the right number of dipoles but still overestimates.
More  experiments under the same setup can be found in the supplementary materials, to support the accuracy of the performance of SPE.

\begin{table}[htbp]
\caption[Comparison of results from PCA, AIC, MDL, EIF and SPE with simulated data]{Comparison of results from PCA, AIC, MDL and EIF and spiked population eigenvalues (SPE) with simulated data. The first column shows the signal-to-noise ratios. The second to eighth columns are the estimated number of dipoles from the simulated data, where the true number of dipoles is four.}
\centering % used for centering table
 \scalebox{0.8}{
\begin{tabular}{c c c c c c c c c} % centered columns (4 columns)
\hline\hline %inserts double horizontal lines
SNR  & PCA (0.9/0.8/0.7) & AIC & MDL & EIF & SPE (FFT) & SPE (RS) & SPE (TH)\\ [0.0ex] % inserts table
%heading
\hline % inserts single horizontal line
 Noise=0 & 3/3/3     & 120-127 & 120-127 & 4-127& 4 & 4 & 4 \\ [0.0ex] % inserts table
 1     &   21/3/2    & 120-127 & 120-127 & 3-19 & 4 & 4 & 4\\ [0.0ex] % inserts table
 .1    &   69/48/34  & 122-127 & 120-127 & 1-11 & 4 & 4 & 4\\ [0.0ex] % inserts table
 .01   &   76/57/44 & 121-127 & 120-127 & 1-11 & 4 & 4 & 4\\ [0.0ex] % inserts table
 .001  &   78/59/46 & 120-127 & 120-127 & 1-10 & 4 & 4 & 4\\ [0.0ex] % inserts table
 .0001 &   78/59/46 & 121-127 & 121-127 & 1-13 & 4 & 4 & 4\\ [0.0ex] % inserts table
 % [1ex] adds vertical space
\hline %inserts single line
\end{tabular}
}
\label{2tb3}
\end{table}

%\begin{sidewaysfigure}[htbp]
\begin{figure}
\centering
\includegraphics[width=1.59in]{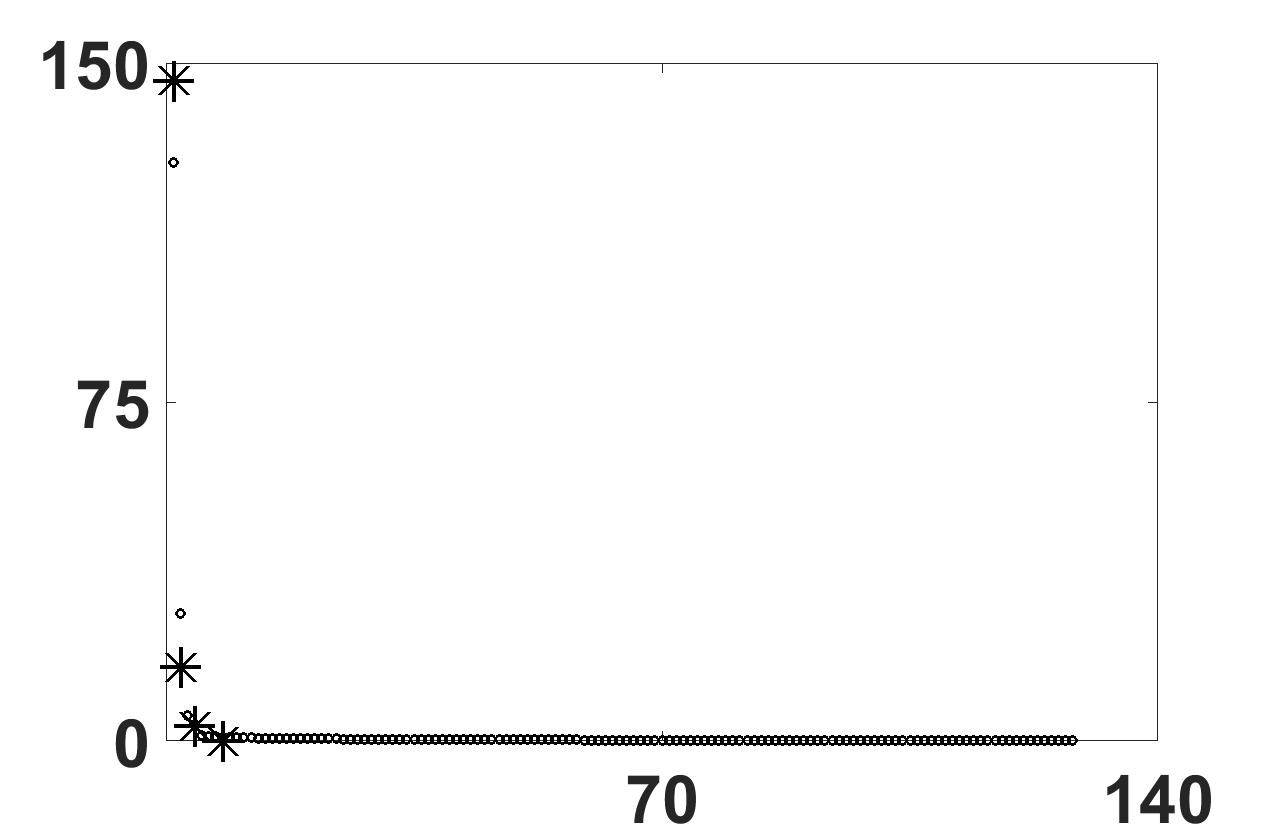}
\includegraphics[width=1.59in]{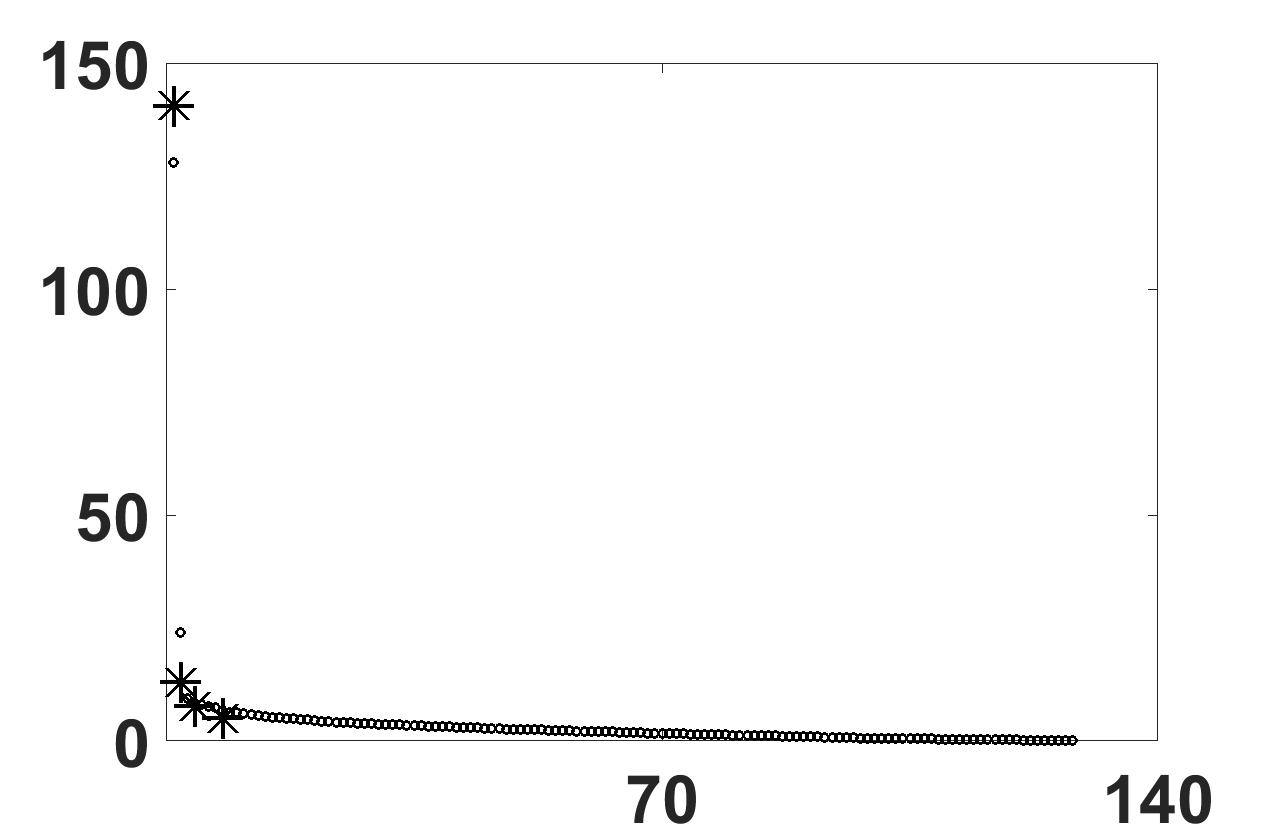}
\includegraphics[width=1.59in]{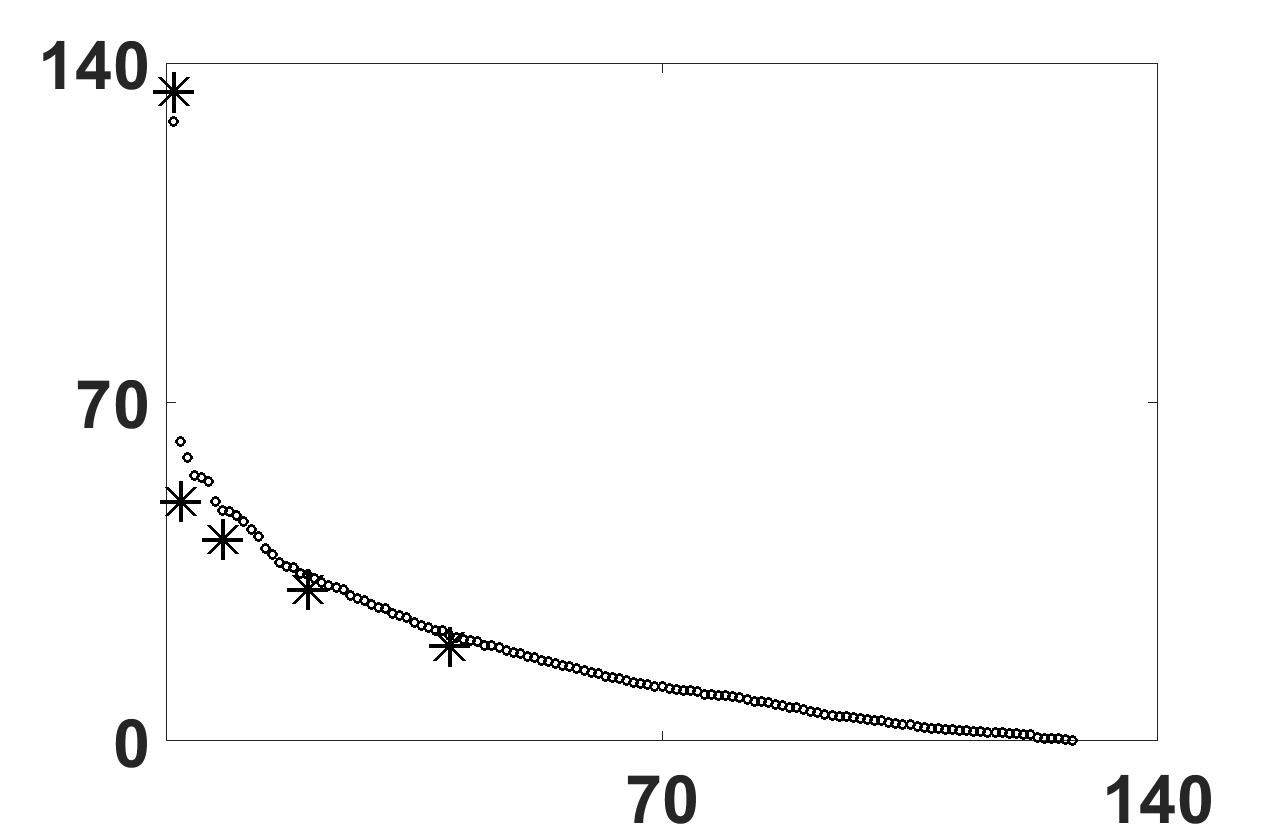}
\includegraphics[width=1.59in]{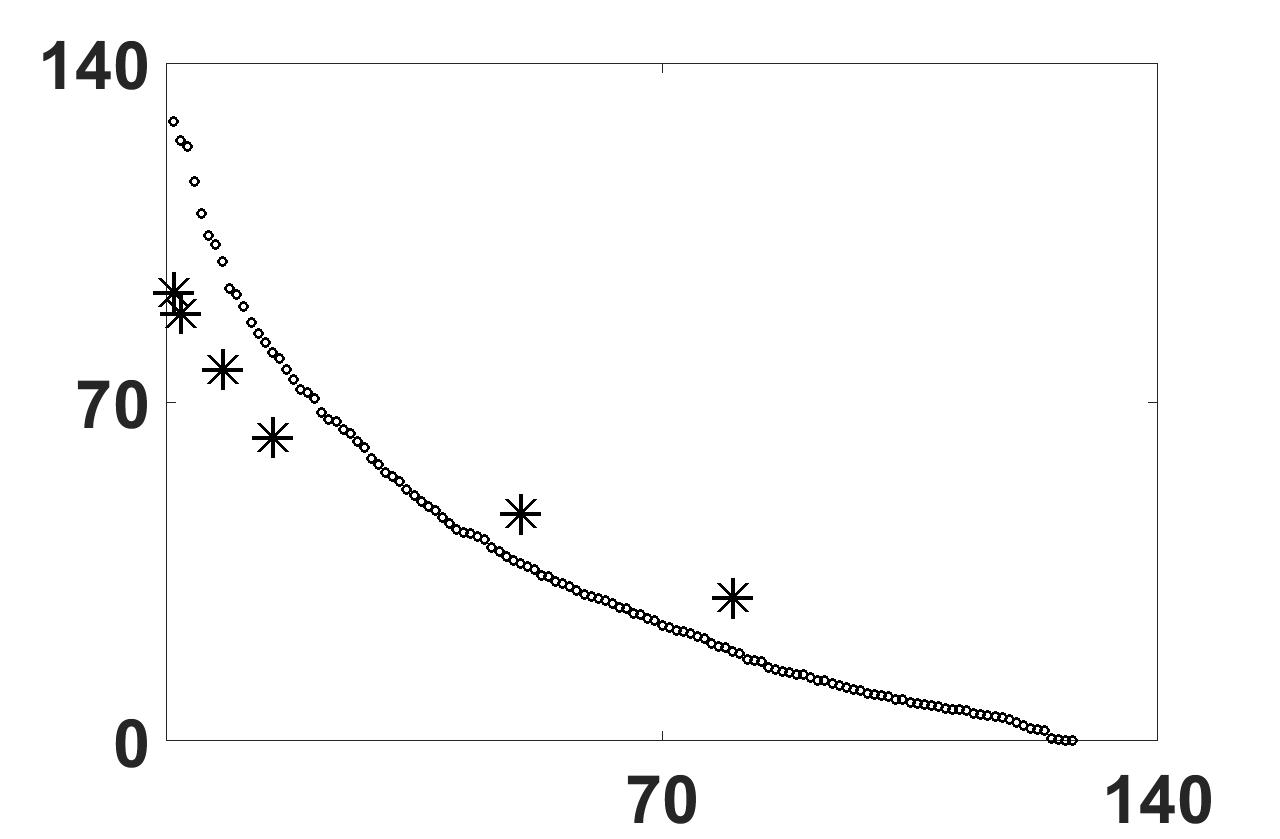}\\
\includegraphics[width=1.59in]{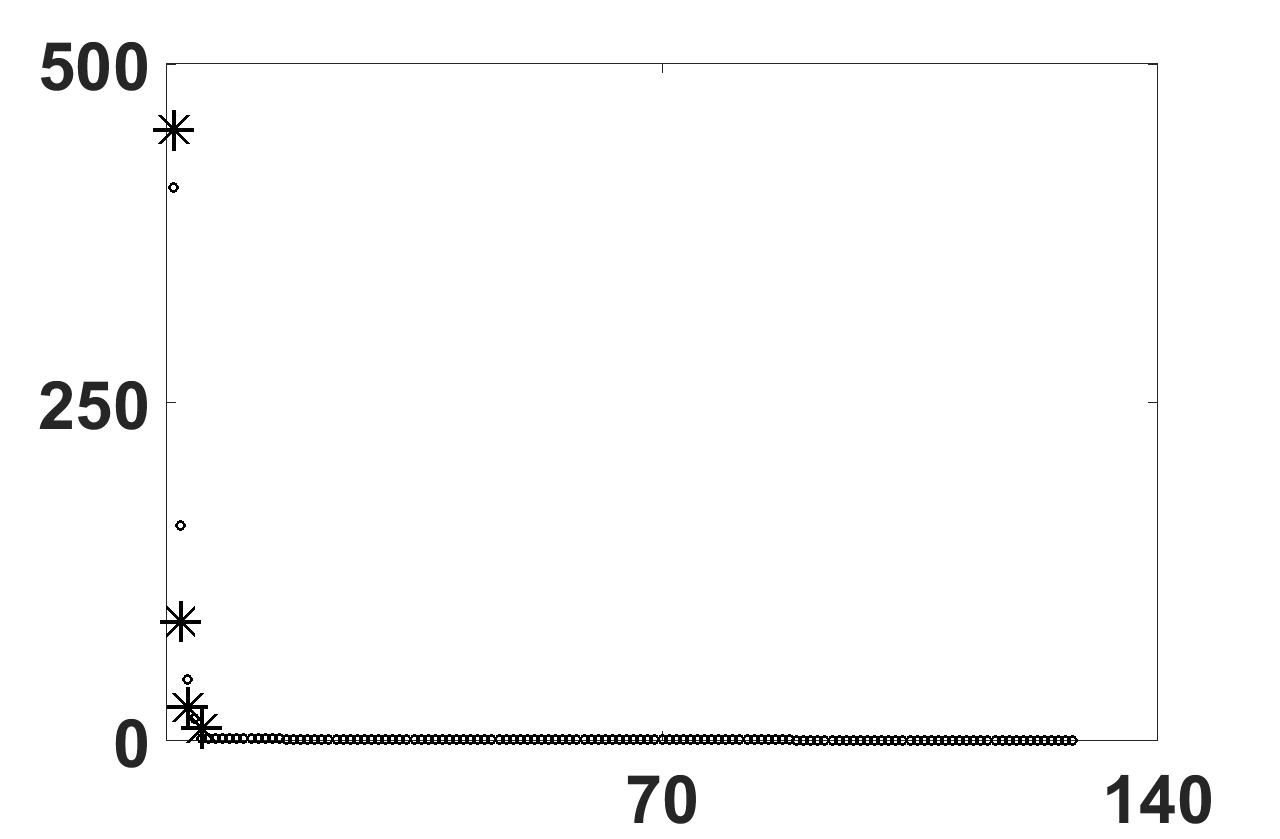}
\includegraphics[width=1.59in]{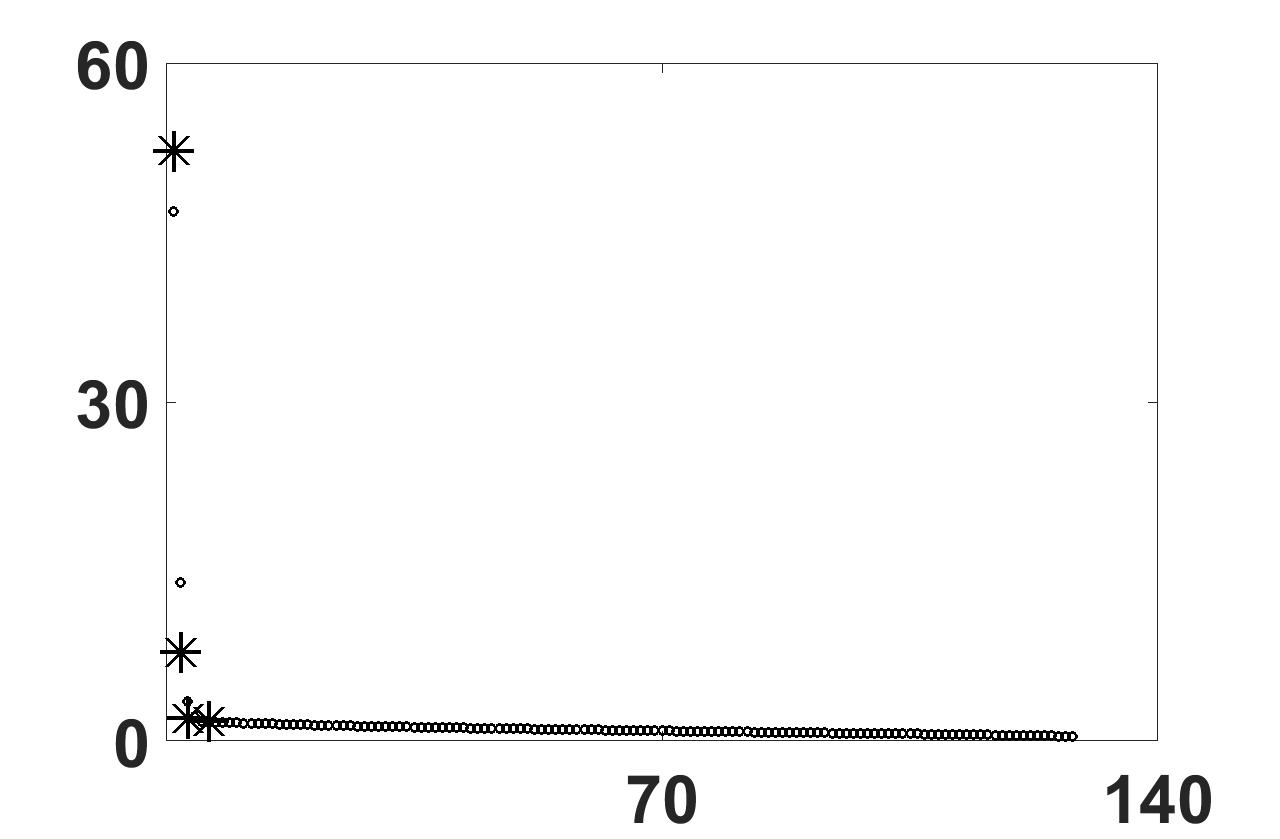}
\includegraphics[width=1.59in]{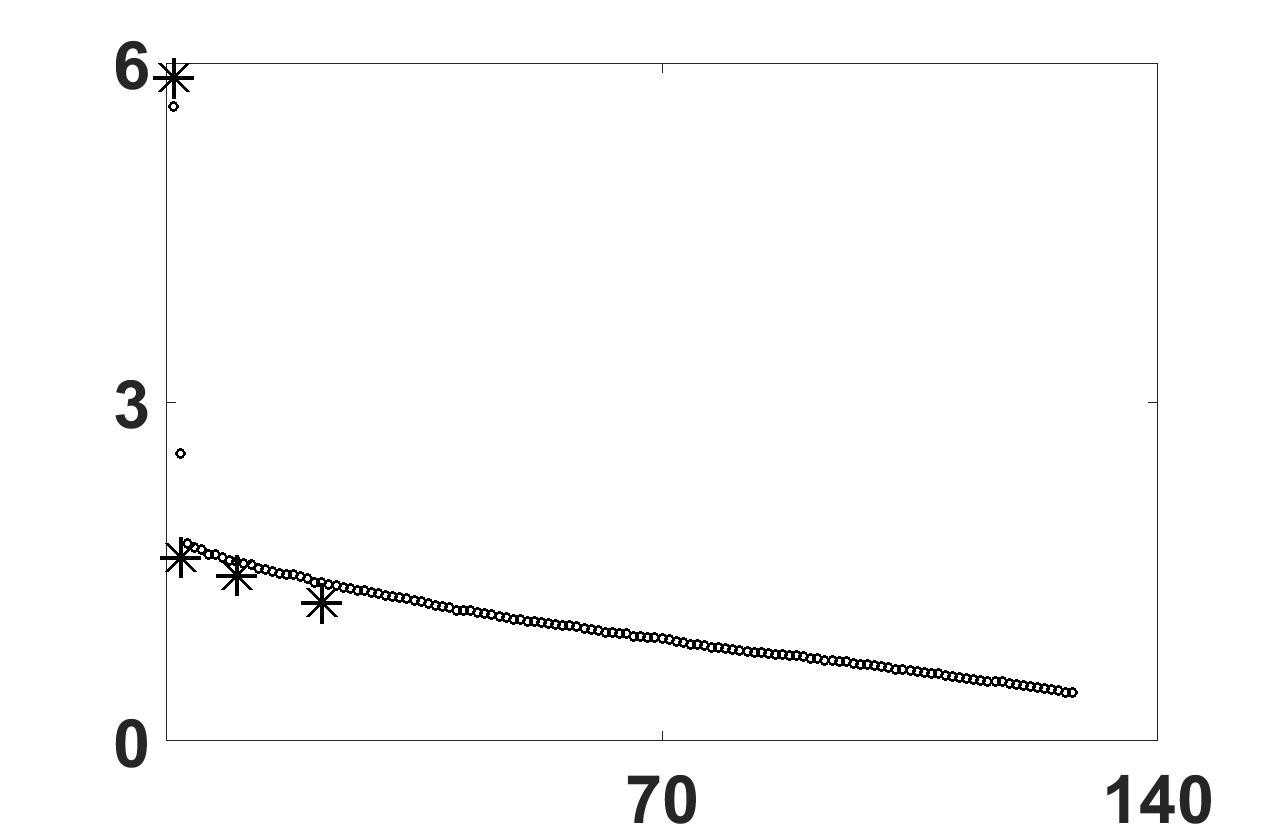}
\includegraphics[width=1.59in]{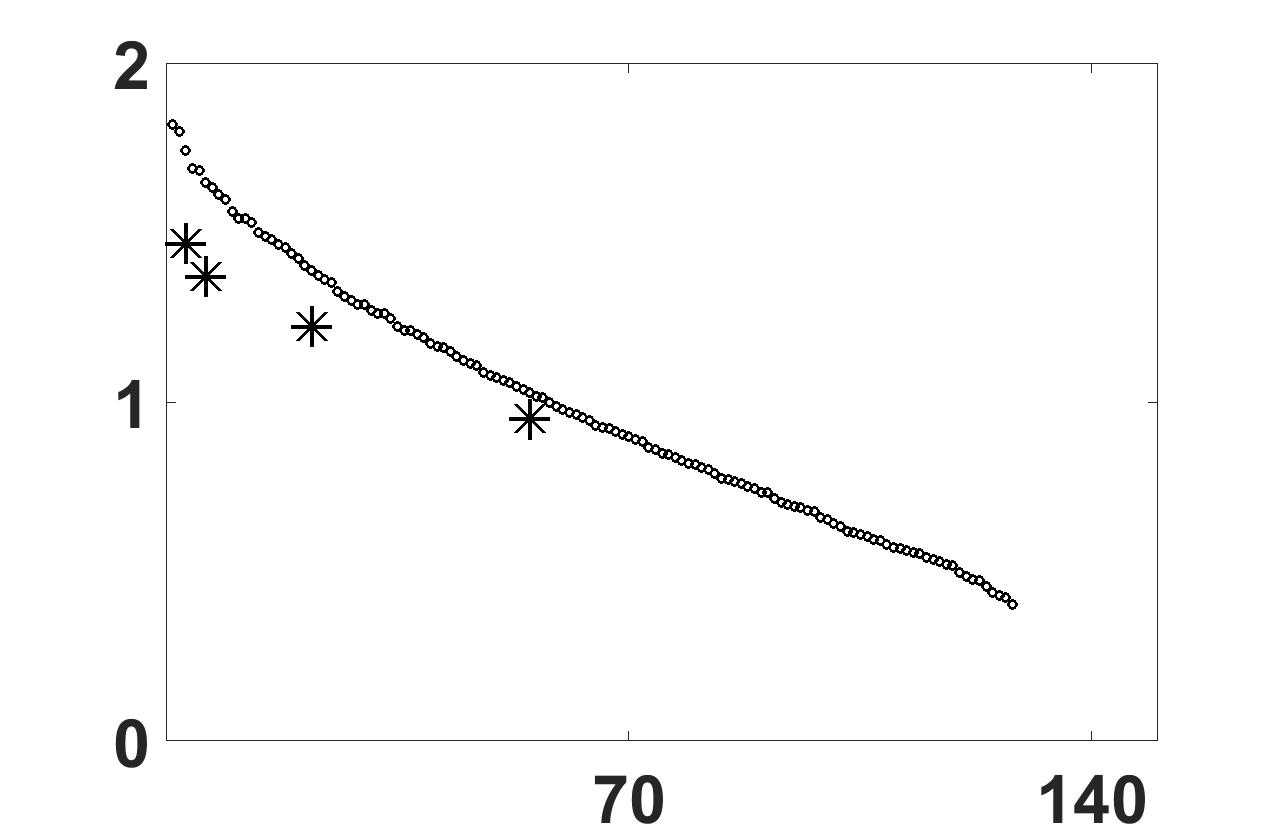}\\
\includegraphics[width=1.59in]{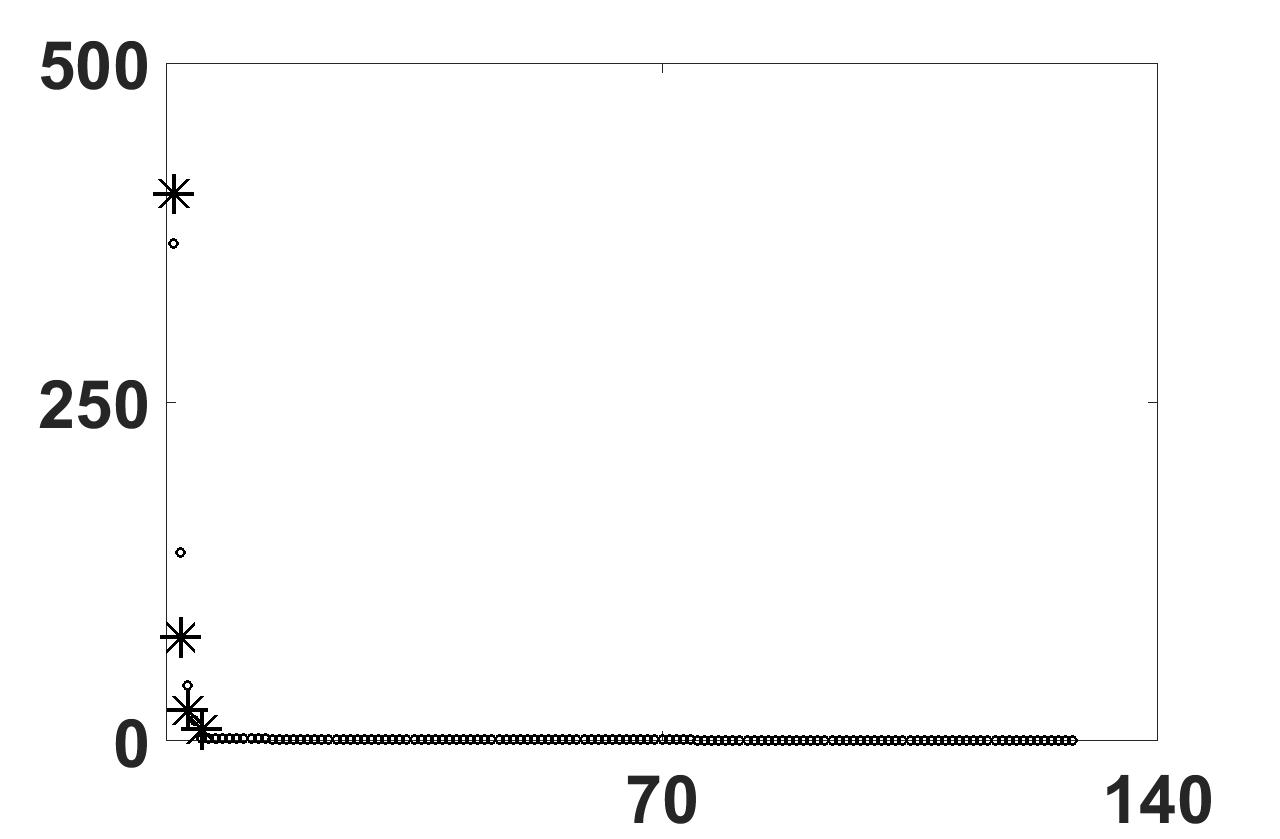}
\includegraphics[width=1.59in]{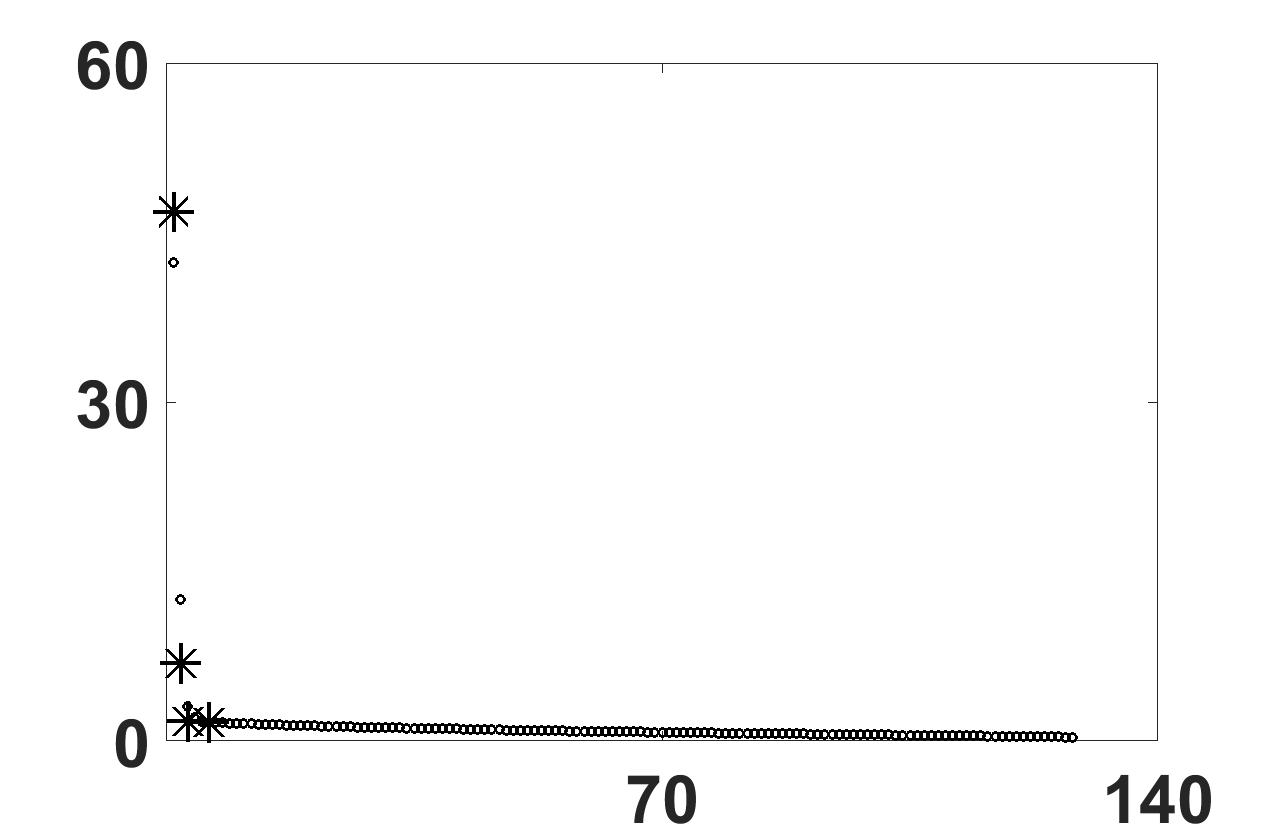}
\includegraphics[width=1.59in]{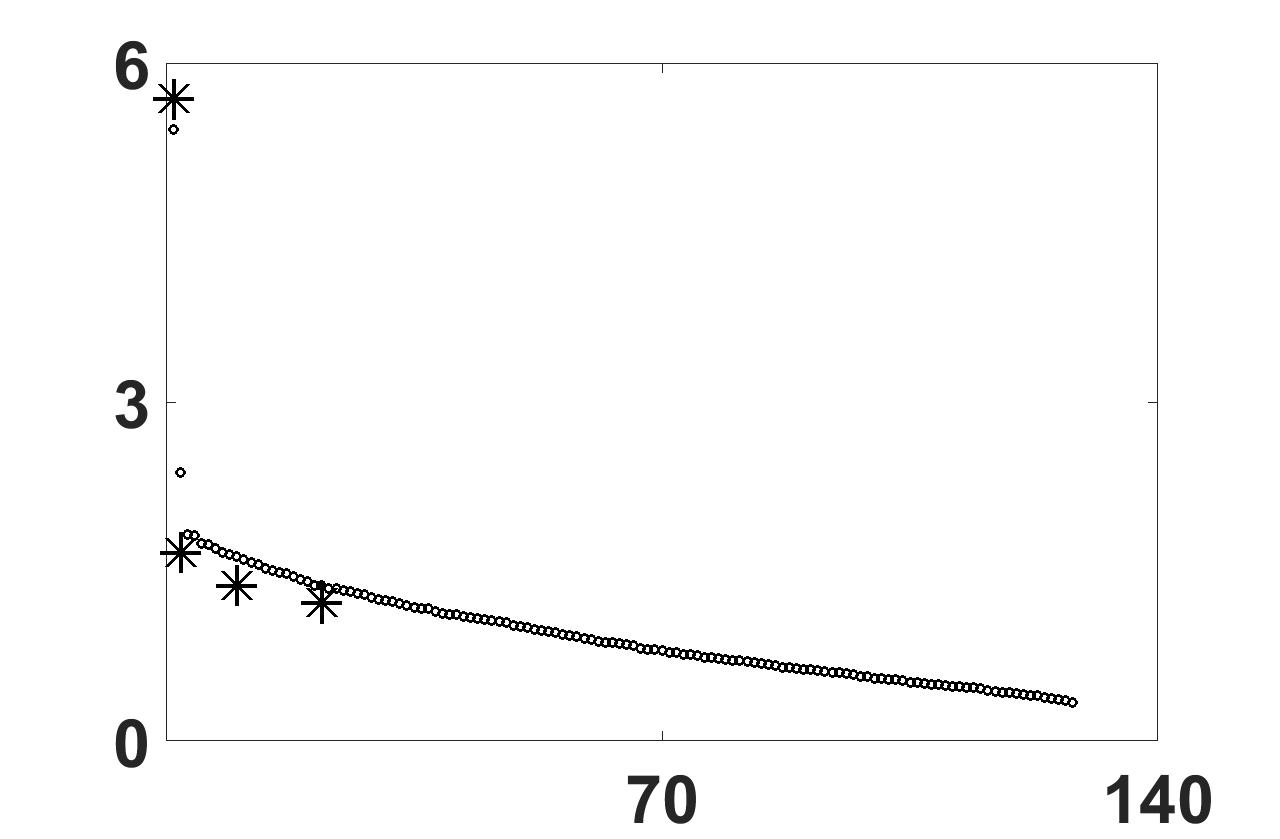}
\includegraphics[width=1.59in]{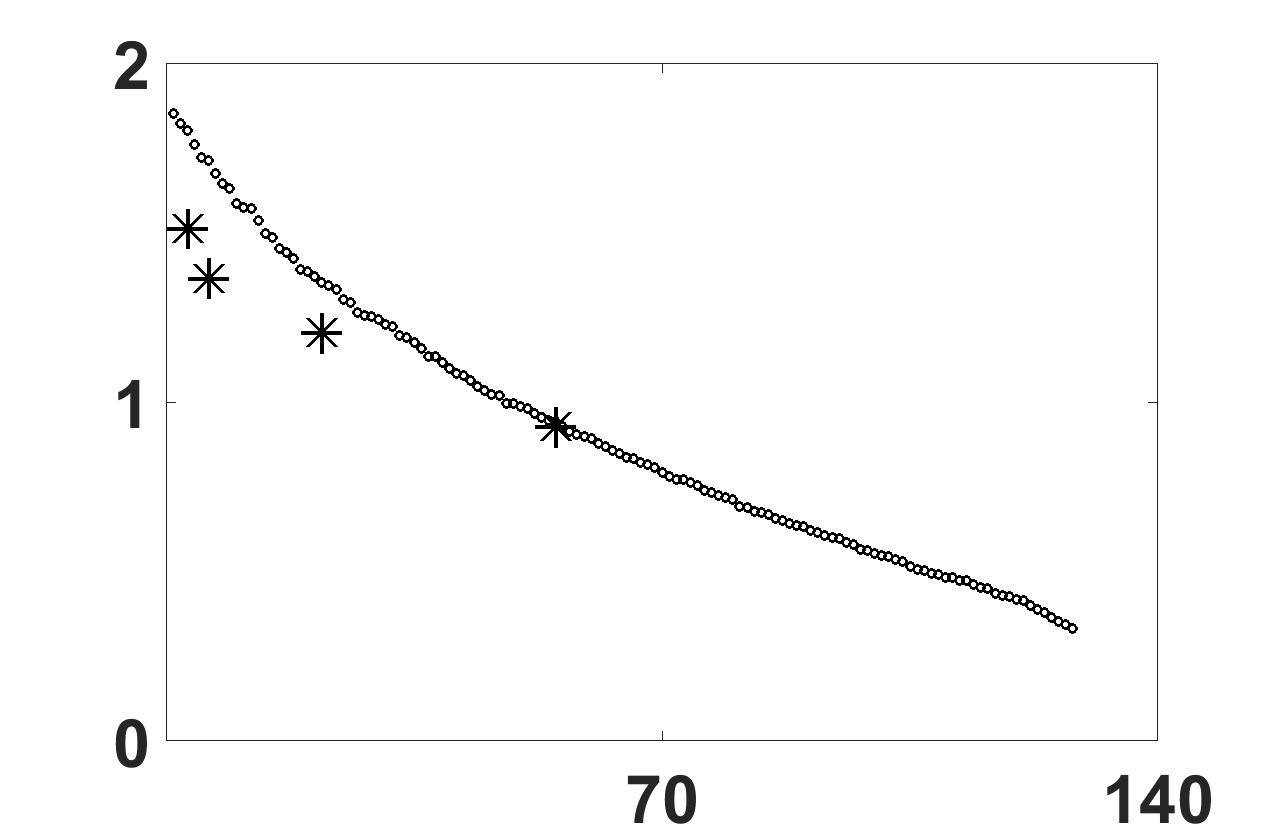}\\
\includegraphics[width=1.59in]{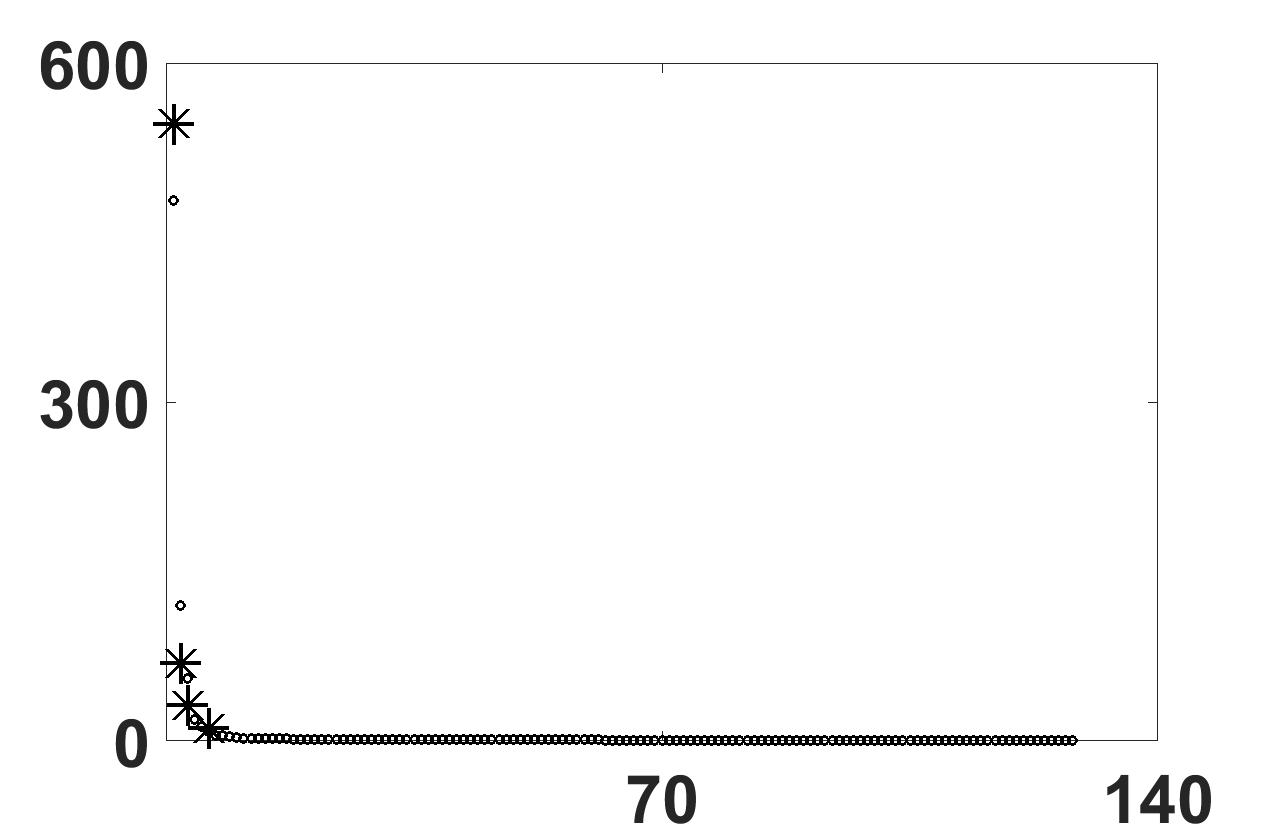}
\includegraphics[width=1.59in]{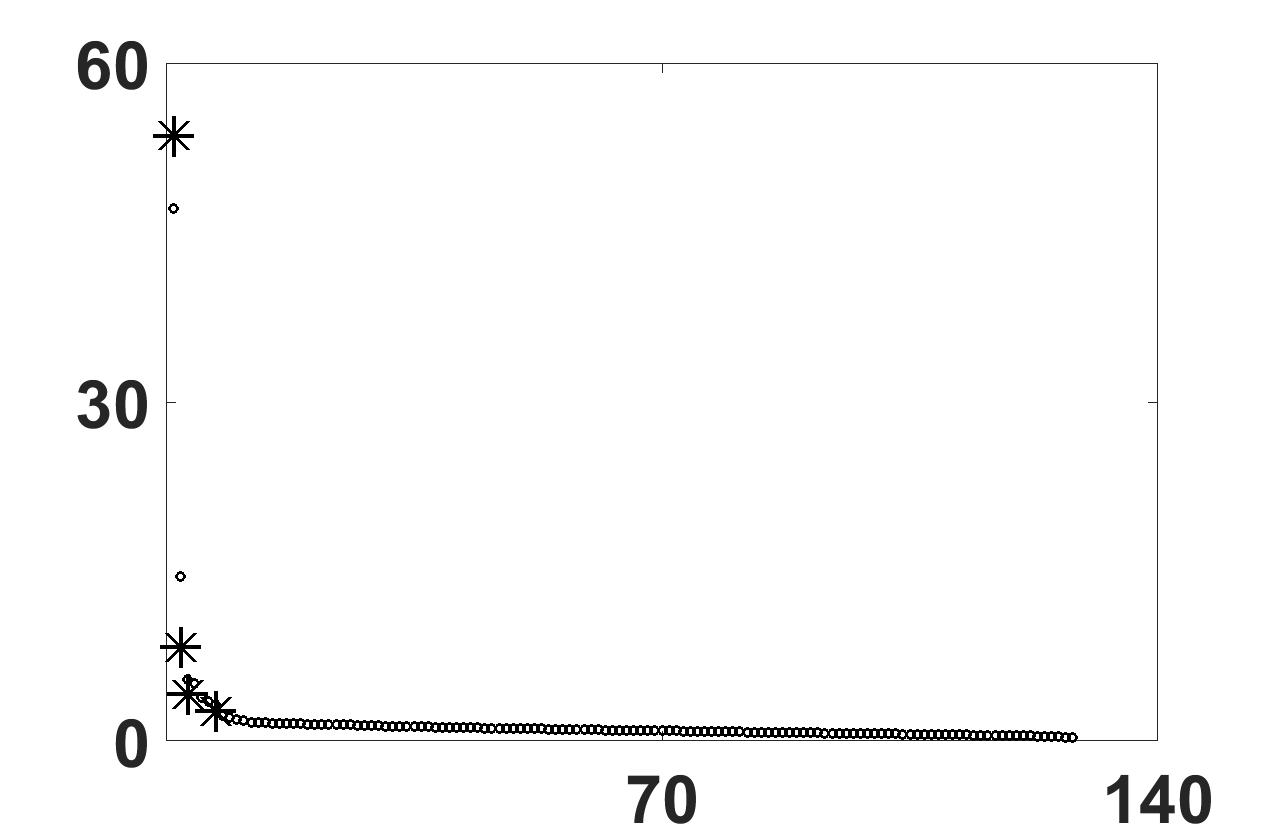}
\includegraphics[width=1.59in]{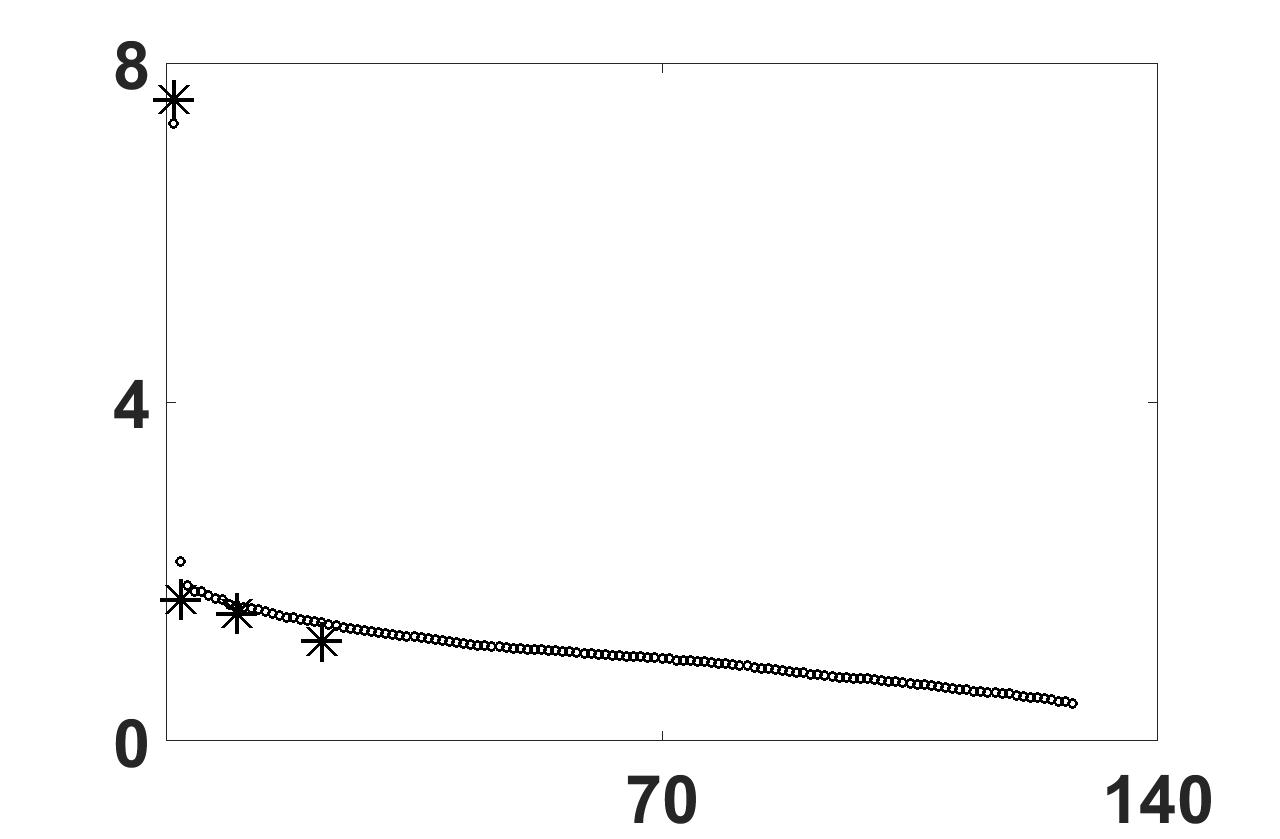}
\includegraphics[width=1.59in]{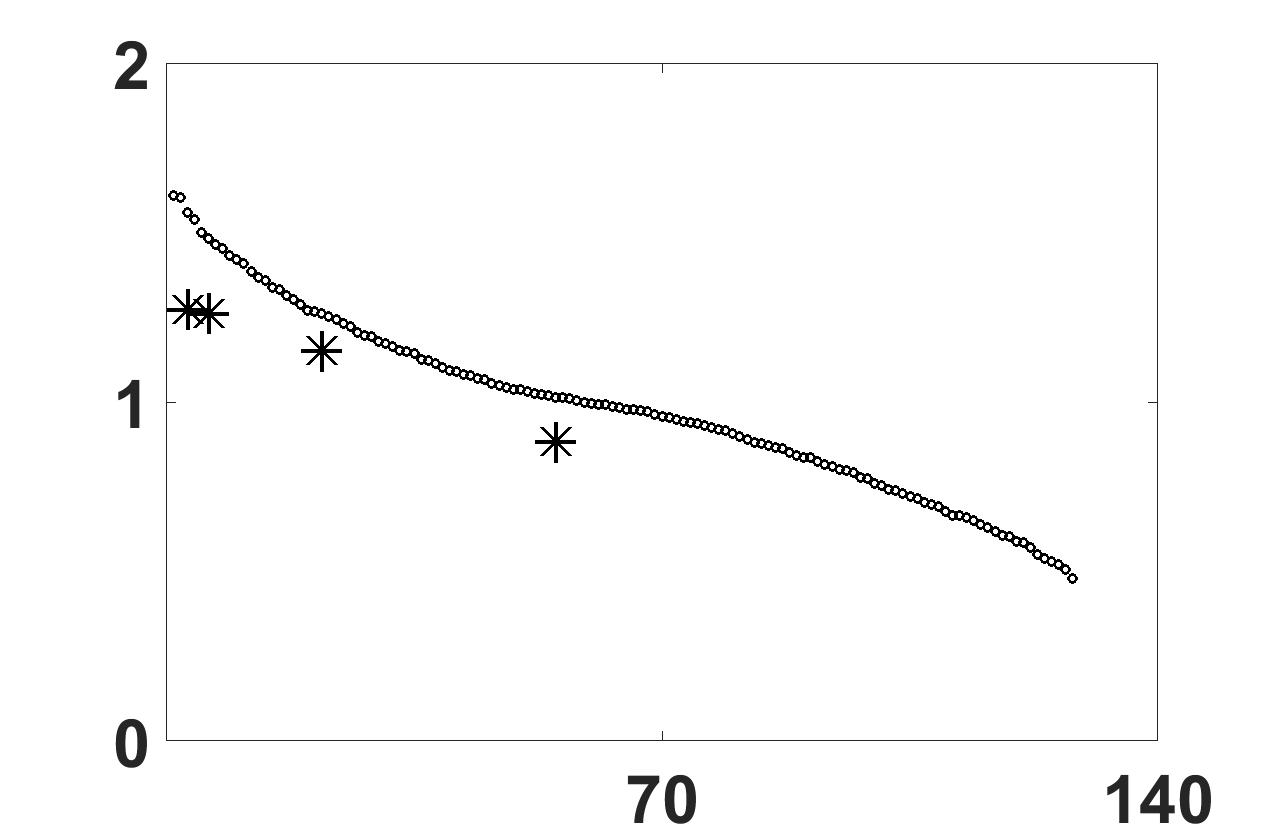}\\
\caption[Illustration of the spiked eigenvalues estimated from the covariance matrix $\widehat{\mathbf{R}}$ and the adjusted covariance matrix $\widehat{\mathbf{R}}_{adj}$]{Illustration of spiked eigenvalues estimated from the covariance matrix $\widehat{\mathbf{R}}$ and the adjusted covariance matrix $\widehat{\mathbf{R}}_{adj}$ at four different SNRs (totally $4\times4$ subplots). Row 1: in each subplot (from left to right), the estimated spiked population eigenvalues are labeled as ``$\ast$'' and the sample eigenvalues are labeled as ``$\circ$'' at SNR of $1$, $.1$, $.01$ and $.001$. The Rows 2-4 give the same information for $\widehat{\mathbf{R}}_{adj}$, estimated respectively by Fourier transform, residual analysis and thresholding method.}
\label{spk}
\end{figure}
%\end{sidewaysfigure}
%\newpage

\subsection{Hunting Unknown Signal in the MEG Room}
\label{empty-room}
For real MEG data on human subjects, we cannot clarify the accuracy of our methods, since the truth of how many sources may be present in the data is unknown. However, it would still be quite interesting to see the performance of our methods on specific MEG data where we do know the truth. In the following analysis, a dataset from an empty MEG room will be used;  that is, there is no subject in the MEG room. To our knowledge, all the devices in the room that might cause electric potential were turned off, but one device was constantly producing energy around 60 Hz. The magnetic field distribution was recorded by a 306-channel system. A small portion of the dataset, 5000 milliseconds long with only 102 channels (magnetometers), was used in our analysis. Conservatively speaking, there was only one source (60 Hz), or at least one, with high frequency in our data. Our attempt was to verify the existence of this high frequency source, and to estimate the number of active sources in the room, using our proposed method on this data. %Those 102 channels were direct magnetic field measurements (the other 204 channels were measuring the change of the magnetic field). 

\begin{figure}[h]
\begin{center}
\subfigure[]{
\includegraphics[width=2.3in]{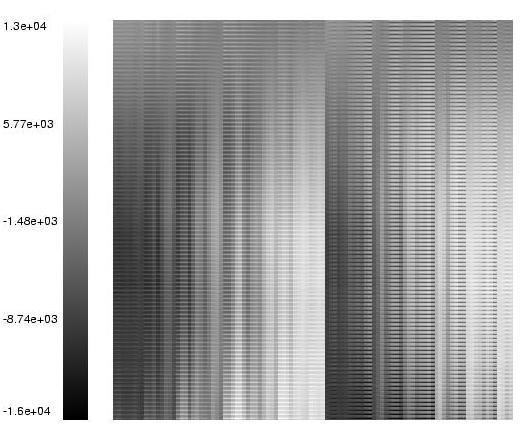}}
%\hspace{.2in}
\subfigure[]{
\includegraphics[width=2.3in]{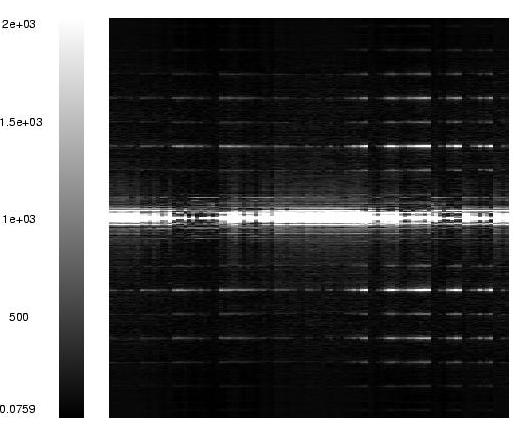}}
\end{center}
\caption[Raw data (empty room) and the modulus plot of the data after the Fourier transform]{Raw data (empty room) and the modulus plot of the data after the Fourier transform. (a) The gray scale plot of the raw data of 5000 milliseconds. The horizontal axis is time (milliseconds); the vertical axis is the channel number (102 channels in total). (b) The modulus plot of the complex-valued Fourier coefficients of the raw data. The horizontal axis is time (milliseconds); the vertical axis is frequency.}
\label{fig2}
\end{figure}

The magnitude of the raw data (Figure \ref{fig2}(a)) in the empty room is in the range of $-1.6\times 10^{4}$ fT to $1.3\times 10^{4}$ fT. We can see that the white lines are equally distant in the modulus plot (Figure \ref{fig2}(b)) of complex Fourier coefficients truncated to 2000 for raw data. This is a clear indication of a periodic source at about 60 Hz in the data. To see if the number of sources that our method detects includes the 60 Hz one, it is necessary for us to run the same analysis in an environment when the 60 Hz is not available. This means we need to filter the 60 Hz signal from the raw data. In fact, we filtered all frequencies above 50Hz. Figure \ref{fig4}(a) shows the modulus plot of the Fourier coefficients after filtering all frequencies above 50Hz; all the white lines associated with 60 Hz, 120 Hz, 180 Hz, and so on, disappear. The image after filtering (shown in Figure \ref{fig4}(b)) is reconstructed by the inverse Fourier transform of the real part after filtering. We do not show the imaginary part of the filtered inverse transformed data, because the figures are all nearly zero (less than $10\times 10^{-11}$ fT).

%We list the 10 largest eigenvalues from each of the four methods. NPW finds there are three eigenvalues above the threshold. But there is a significantly large eigenvalue out of the three and it is significantly greater than the second one; the second one is significantly greater than the third, so the threshold does not matter too much here. Thus, we report by NPW one or two sources exist in the data (Figure \ref{fig3}(a)). Similarly, we report three dipoles by NPR (Figure \ref{fig3}(b)) and three dipoles by NPF (Figure \ref{fig3}(c)). PCA finds two or three significant sources from the data (Figure \ref{fig3}(d)).

Note that the values of the measurements are very small in the empty room, and this is particularly true for the data after filtering. To avoid possible round-off error due to floating-point arithmetic, we proceed to normalize the original covariance matrix $\widehat{\mathbf{R}}$ by
\begin{equation*}
\widehat{\mathbf{R}}^{*} = \phi(K) \cdot \frac{\widehat{\mathbf{R}}}{\|\widehat{\mathbf{R}}\|_2}
\end{equation*}
where $K$ is the number of sensors ($K=128$ in this experiment) and function $\phi$ describes the contribution of the dimension to the problem. In practice, one can choose $\phi(K)=1,K,K^2$ (we set $\phi(K)=K$ here).

{\bf Remark:}  (1) For the noise estimation, we use the corresponding normalized data by $\mathbf{Y}_{\tiny\mbox{norm}}(t)= \sqrt{K/\|\widehat{\mathbf{R}}\|_2} \cdot\mathbf{Y}(t)$. (2) When the noise effect is negligible, no optimal SNR transformation is needed since a bad noise estimation may even deteriorate the accuracy of the result.

%{\bf Remark}  when the noise effect is negligible, no optimal SNR transformation is needed. Moreover, a bad noise estimation may even deteriorate the result. Therefore, in the case of data after filtering, instead of normalizing $\mathbf{R}_{adj}$ we normalize the original matrix $\mathbf{R}$ itself.
%(1) There is no need to do unit transformation of the original data since the normalization procedure will cure the round-off error automatically. (2) the threshold selection method of $\epsilon$ depends on the dimension contribution function $\phi(K)$.

\begin{figure}[h]
\begin{center}
\subfigure[]{
\includegraphics[width=2.3in]{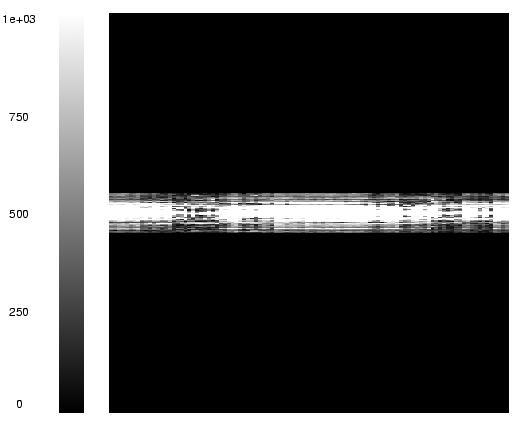}}
%\hspace{.2in}
\subfigure[]{
\includegraphics[width=2.3in]{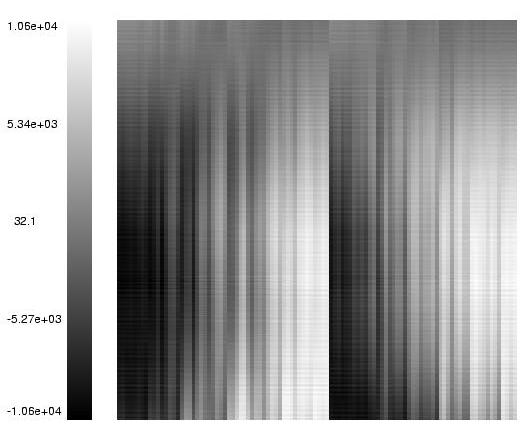}}
\end{center}
%\footnotesize{Figure 6: PVM Structure (Master and Workers' jobs)}
\caption[Raw data of the empty room after filtering]{Raw data of the empty room after filtering. (a) The modulus plot of the complex-valued Fourier coefficients after filtering all frequencies above 50 Hz truncated to 1000. The horizontal axis is time (milliseconds); the vertical axis is frequency. (b) The gray scale plot of the real part of inverse Fourier transform after filtering coefficients; that is, the data after all coefficients above 50 Hz are zeroed out.}
\label{fig4}
\end{figure}

A summary of the performance of the different methods (PCA, AIC, MDL, EIF and SPE) applied on this data (before filtering and after filtering) is shown in Table  \ref{3tb3}. Both the AIC and MDL underestimate the number of signal sources, while the EIF overestimates the number of signal sources. All are not able to tell the difference in the number of sources before and after filtering. Both PCA and SPE provide a reasonable estimate of the number of sources: four before filtering and three after filtering for PCA, three before filtering and two after filtering for SPE, while PCA tends to pick up more sources. This leads us to believe that there is at least one (or even two) other active sources ($<$ 50 Hz) that exist in the MEG room.

\begin{table}[!htb]
\caption[Comparison of results from PCA, AIC, MDL, EIF and SPE with the raw data of the empty room]{Comparison of results from PCA, AIC, MDL, EIF and SPE with the raw data of the empty room.}
\centering
\begin{tabular}{c c c c c c} % centered columns (4 columns)
\hline\hline %inserts double horizontal lines
  & PCA & AIC & MDL & EIF & SPE (FFT/RS/TH) \\ [0.0ex] % inserts table
\hline % inserts single horizontal line
Before filtering & 4 & 1 & 1 & 2-31 & 3 \\ [0.0ex] % inserts table
After filtering  & 3 & 1-2 & 1-2 & 3-30 & 2 \\ [0.0ex] % inserts table
\hline %inserts single line
\end{tabular}
\label{3tb3}
\end{table}

\section{Brain-controlled interfaces Data}
\label{realdata}
The real data analysis reports results of finding varying brain sources in a Brain-Controlled Interfaces (BCI) experiment. The data consists of 28000-millisecond recordings collected at the Center for Advanced Brain Magnetic Source Imaging (CABMSI) at Presbyterian University Hospital in Pittsburgh. In the first part of the experiment, the subjects were asked to imagine performing the ``center-out'' task using the wrist (imagined movement task), and in the second part, the subjects controlled a 2-D cursor using the wrist to perform the center-out task following a visual target (overt movement task). The subject, as illustrated in  Figure \ref{experiment} (the time scale in the real experiment can be different), holds the 2-D cursor and waits for the cursor to go to the center. To complete the trial successfully, the cursor stays at the center for a short period until the peripheral target appears. The cursor needs to move out to the target and stays there for another short period. The target changes color when hit by the cursor, and disappears when the holding period has finished.  %(see more explanation in \cite{Wei_2010}).

\begin{figure}[ht]
%\begin{figure}
\centering
\includegraphics[width=3.0in]{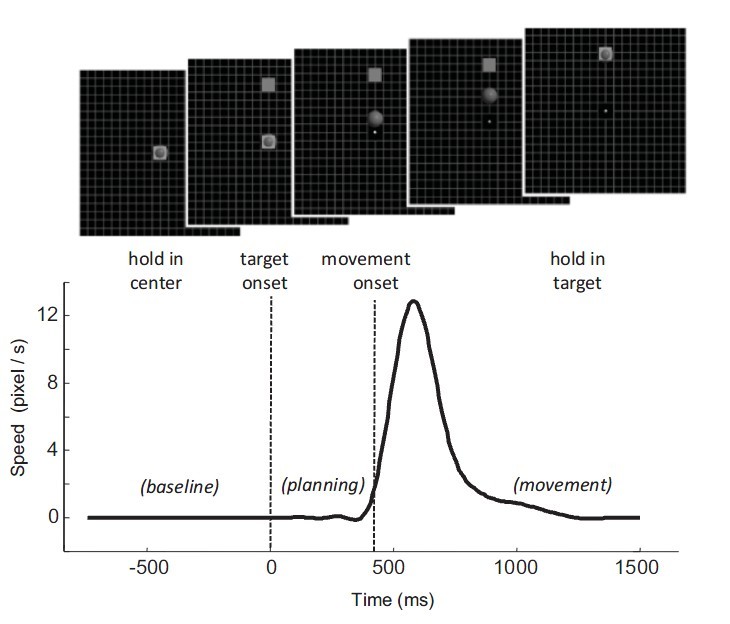}
\caption[A BCI experiment chart]{A BCI experiment chart. The bottom trace shows the speed profile of the cursor from a representative trial, and the dotted lines delimit the pre-movement/planning period. }
%\end{figure}
\label{experiment}
\end{figure}

\begin{figure}[ht]
%\begin{figure}
\centering
\subfigure[]{
\includegraphics[clip, trim=0 3.3in 0 3.3in, width=2.5in]{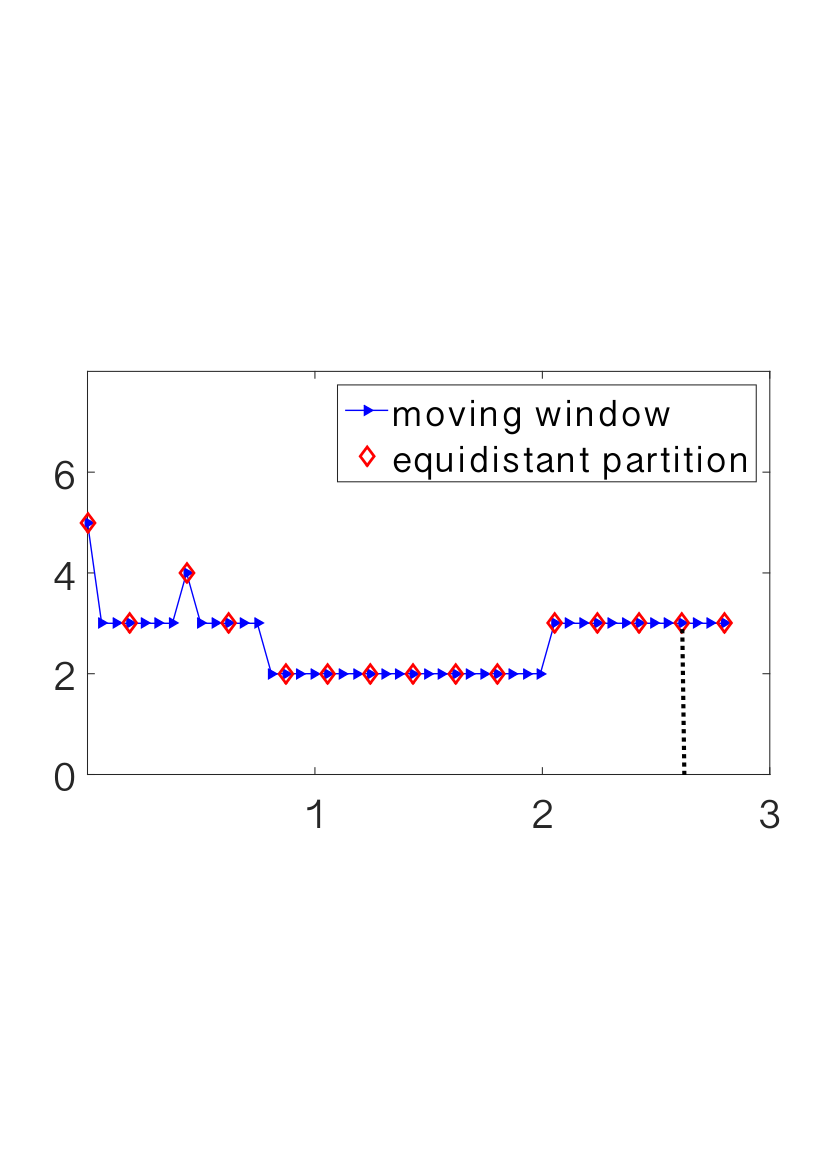}}
\subfigure[]{
\includegraphics[clip, trim=0 3.3in 0 3.3in, width=2.5in]{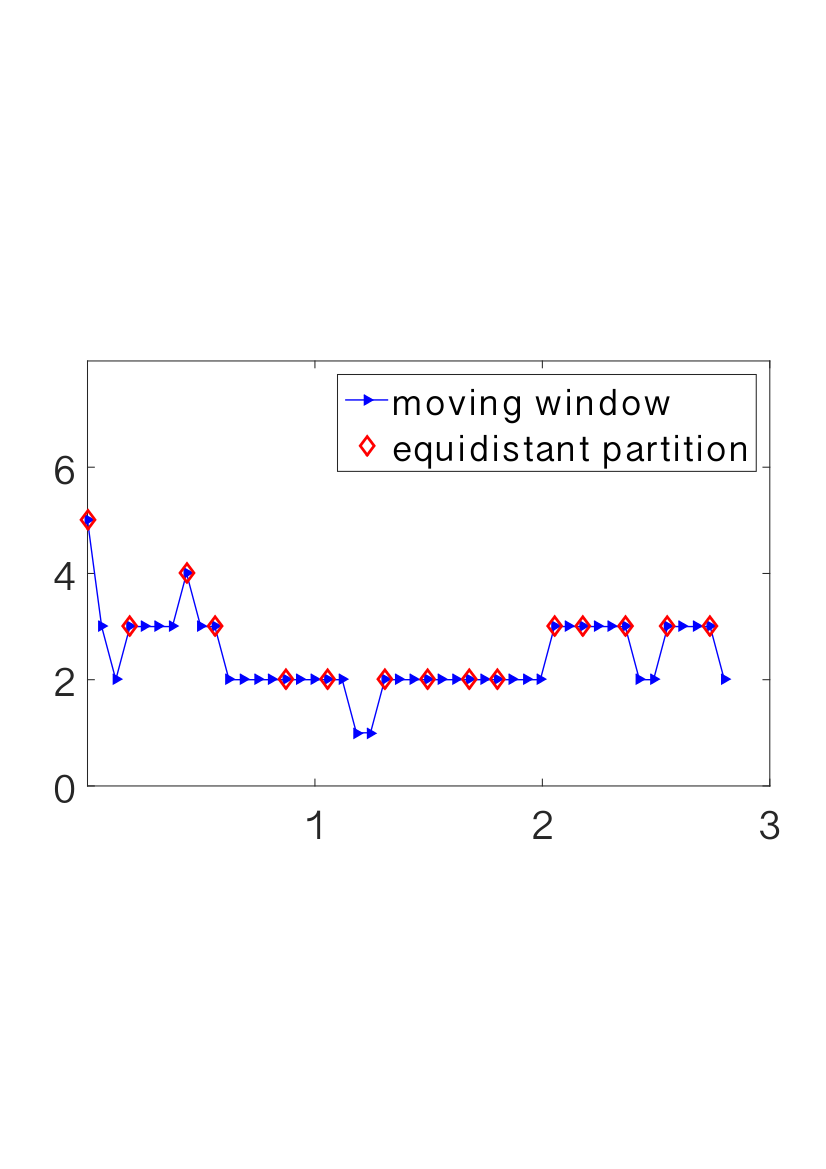}}
\subfigure[]{
\includegraphics[clip, trim=0 3.3in 0 3.3in, width=2.5in]{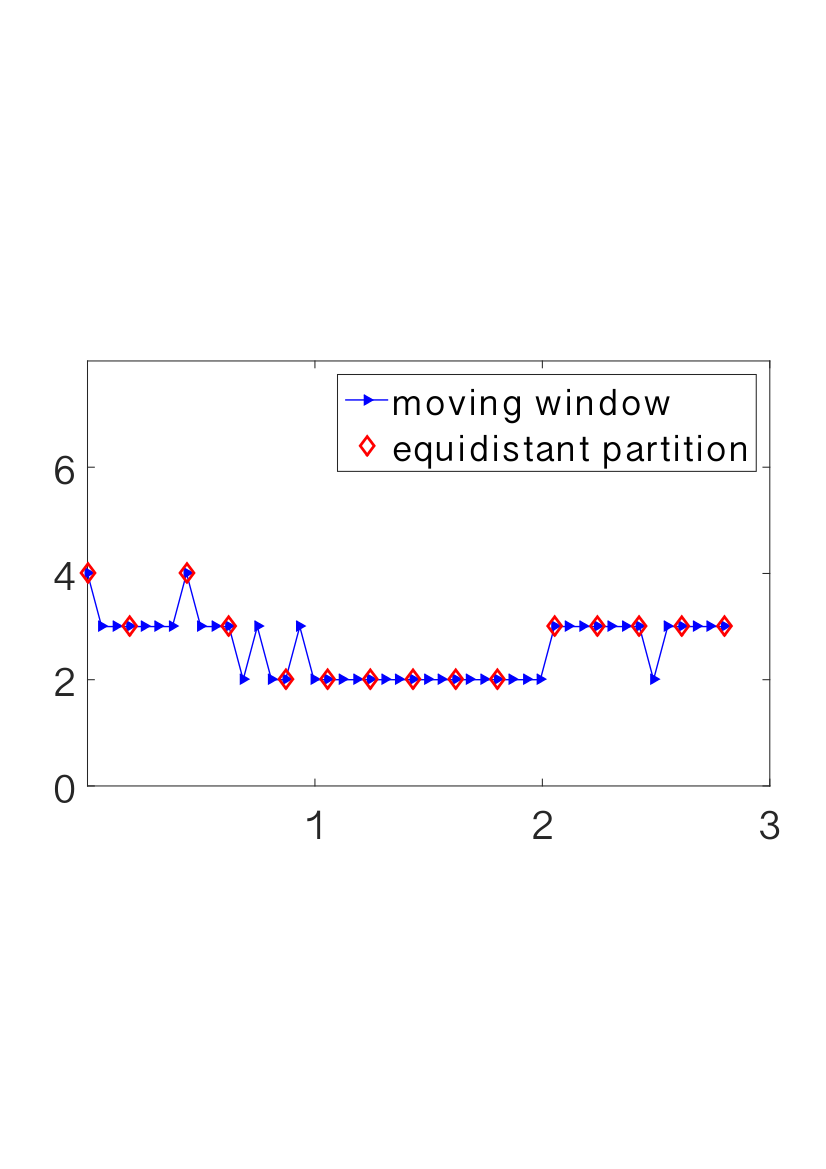}}
\subfigure[]{
\includegraphics[clip, trim=0 3.3in 0 3.3in, width=2.5in]{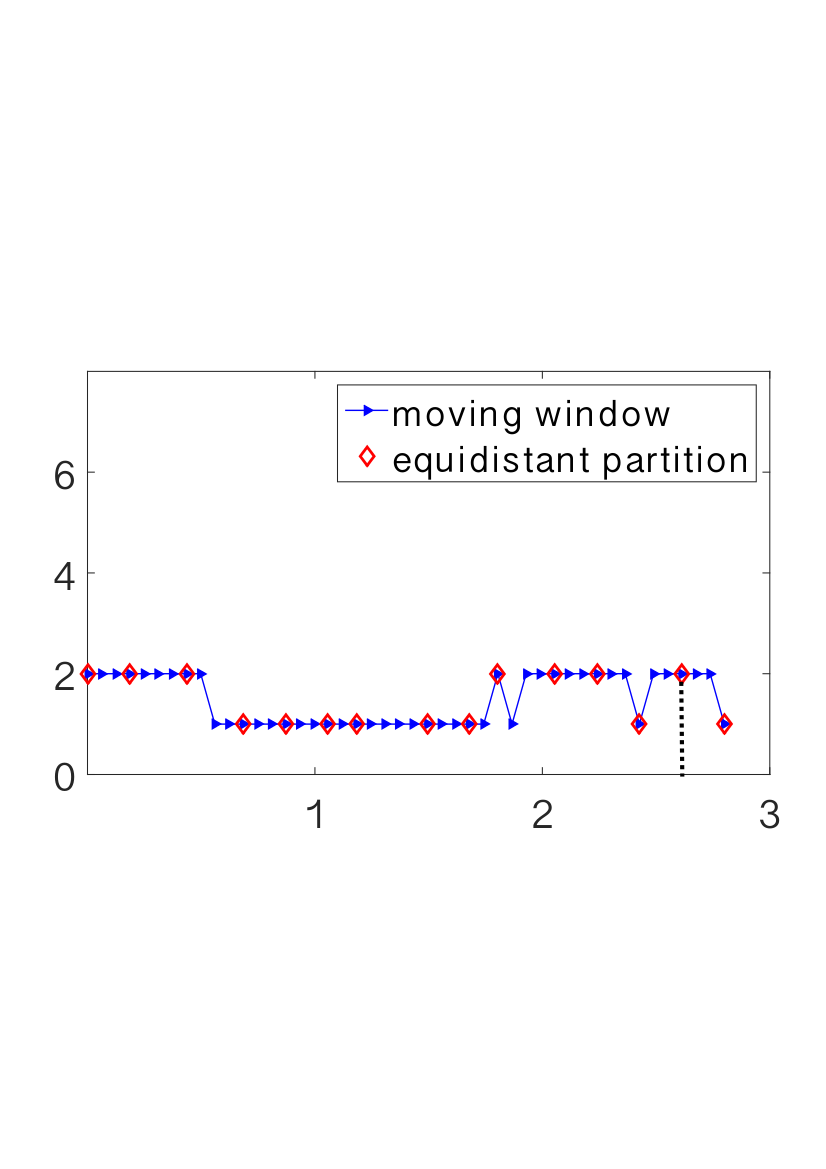}}
\caption[Estimation of the number of signal sources in the BCI data]{Estimation of the number of signal sources in the BCI data. The horizontal axis is time (seconds); the vertical axis is number of estimated sources. a) the noise estimation is based on the Fourier transform; the blue curve shows the estimated number of sources for each moving window with a length of 2000 ms; the estimated number of sources at each equidistant partition (2000 ms) of the data is also shown as a diamond in red. (b) and (c) display the same information, but with the noise estimated by the residual analysis and thresholding methods, respectively. (d) displays the same result with no noise estimation.}
\label{source-map}
\end{figure}

The analysis is based on the raw MEG data from 102 magnetometers, without performing spatial filtering or smoothing. The goal is to investigate the dynamics of the sources in the data. Figure \ref{source-map} presents the evolution of the number of possible active sources estimated during the experiment. As can be expected, there tend to be more active sources during the imagined movement task period than overt movement task period. This reveals a possible delay of the observed magnetic signal reaching its peak. This can also be partially explained by the fact that the subject was engaged in the imagined movement task by the experimenter, i.e., catch trials were inserted.  The data also serves to show that there are still some active sources presented after the movement, which might come from the holding period of the subject preparing for the next task. 

To illustrate the effect of partition of the data on the estimation, a dual-scale point of view is used: 1) the equidistant partition of the data,  2) a moving window scheme. For the equidistant partition, we equally divide the data into several chunks, each with a  length of 2000 ms; for the moving window, we choose the window of the same length (2000 ms) and move the window from 0 ms at the rate of 600 ms per move. The numbers of the sources estimated by these two methods are depicted in Figure \ref{source-map}, respectively. The effects of choosing which partition to use are not so apparent in this particular data. In this case, we can see that there is no significant change to the estimated sources by moving the window, suggesting that the estimated number of sources at each separate time interval seems good enough to represent the dynamics of the number of sources. This provides us with some flexibility in using the dual-scale point of view when presenting the number of sources. The difference with respect to the number of sources lying in-between the overt and imagined movements shown in our analysis has been consistent with that in \ctn**{Wei_2010}, where the main finding is to decode the intended movement direction in the absence of overt movement using MEG.

We further examine the effect of noise estimation on the number of sources. Interestingly, Figure \ref{source-map}(a), (b) and (c) show that noise estimation has generally detected additional active sources rather than no noise estimation (Figure \ref{source-map}(d)). Since the data has been normalized beforehand, any change of the number of the estimated spiked eigenvalues would imply a change of the sources. This phenomenon is expected as there might be some hidden eigenvalues from the signal part that may not be easily revealed when the noise presents. We note that, during the time period of 24000-25999 ms, the eigenvalues before and after noise estimation (i.e., Fourier transform) are $(102, 3.5, .8, .1, \cdots)$ (see dotted line in Figure \ref{source-map}(a)) and $(102, 10.5, 3.3, .4, \cdots)$ (see dotted line in Figure \ref{source-map}(d)), which suggests that the eigenvalue $10.5$ is possibly from a potential source. Similar conclusions can be drawn for the other two methods of noise estimation. This further indicates that the optimal SNR transformation is helpful in estimating the number of sources.

%\section{DISCUSSION}
\section{Discussion}
\label{2discussion}
The determination of the number of signal sources in the MEG data is a very challenging problem. Due to the noisy characteristics of the MEG data, effective methods for dealing with this problem are lacking. Conventional approaches such as PCA-based methods or methods involving information criteria are essentially not helpful in deciding the dimensionality of the data. The difficulty lies in the fact that those approaches simply use the sample eigenvalue distribution, where the sample eigenvalues are not consistent estimators of the population counterparts when the dimension of the data is proportional to the sample size. As is the case with the MEG data, the sample eigenvalues are still  mixtures of the signal and noise in the data, and it is quite hard to detect the energy that such a signal contributes to the eigenvalues, compared with the noise.

With the aim of proposing a framework that allows the flexible estimation of the intrinsic dimensionality of the data in MEG, we introduced the notation of spiked covariance model. We showed the importance of estimating the eigenvalues of the population covariance matrix, and the difference if this method from only utilizing the sample counterparts. The spiked eigenvalues were found to be interpretable as a measure of the number of signal sources in the MEG data. Depending on how the SNR varies, the spiked covariance model was seen to be more reliable in estimating the number of signal sources. In this sense, it can be thought of as a significant improvement on the methods which utilize the sample eigenvalue distribution.

We show that the optimality of the eigenvalue distribution is achieved on the transformed data matrix rather than the original data matrix. However, the optimal SNR transformation requires a reasonably acceptable accuracy of the noise estimation.  There have always been difficulties in defining noise in neurological experiments; in particular, for the MEG data, the noise structure is very complicated. This paper is mainly focused on two noise structures: the independent noise (Fourier method), and the correlated noise (residual method, thresholding method), where in both situations, we aim to recover the main diagonal of the noise covariance matrix at the sensors. In the examples we considered, the estimated covariance matrices under the optimal SNR yielded visually apparent population eigenvalues that were remarkably different from the sample eigenvalues. From case to case, both methods seemed to capture particularly relevant underlying influential eigenvalues of the data, while the sample eigenvalues were seen as quite noisy. In fact, we noted that the spiked eigenvalues estimated from either method did not have much of an effect on the actual number of the spiked eigenvalues. This points to the potential for future work, on a further investigation of the noise estimation and its influence on the estimated spiked eigenvalues.

The performance of different methods on estimating the number of signal sources was examined in a simulated example, with varying SNRs. In all cases, the spiked population eigenvalues estimated gave quite robust results consistent with the true number of the sources; based on the same transformed covariance data matrix, the PCA as well as AIC, MDL and EIF approaches only worked reasonably well for just a few cases, when the SNR was large, and failed for most situations.  We also attempted to hunt for unknown sources existing in an shielded MEG room. We confirmed the existence of a single 60 Hz source in the room. In addition, another one or two potential sources were detected. One advantage of using our method is that we could possibly detect the hidden signal sources that are different in frequency, which can be further used in identifying high frequency oscillation in the evaluation of epilepsy or other presurgical operations.  In the BCI data,  because the number of sources may change over time, the associated data covariance in the model would change as well. This fact, together the necessity of noise estimation incorporated in the optimal SNR transformation, suggests that a dynamic implementation of the estimation is preferable. We have shown that it is not necessary to estimate the number of sources from the entire data at once, but rather that attempting to estimate it for each time point sequentially can also produce very stable results.

In conclusion, we have been trying to find a way of estimating the number of signal sources in the MEG data. Though the number of spiked population eigenvalues appears to be a useful means of guiding the practitioners' choice on the sources before further applying localization methods, it is certainly not the only choice. The issue of assessing which criterion is the most informative one to decide on the intrinsic dimensionality of the data deserves further scrutiny.  Our method can serve as a reference.
%  and it seems that when new results of the noise estimation is ready,

%\subsection*{Acknowledgments}
%We thank Dr. Anto Bagic for collecting the empty room MEG data and Prof. Rob Kass for the BCI data. Our warm thanks also go to three referees and the associate editor for a number of constructive comments.

%\appendix
%\section*{Appendix A}

\subsection*{Supplementary materials}

A brief description of the forward and inverse problems in MEG, a complete integration of Algorithms 1~-- 3 in Section 2.2, and detailed descriptions of the three algorithms for noise estimation can be found in the online supplementary materials.

\appendix

\section*{Appendix 1: Proof of Theorem 1}\label{thm:snr}
\begin{pf}

For any fixed $\mathbf{X}\in \mathbb{R}^{K \times K}$, there exists a positive number $C$ such that $\|\mathbf{X}\|_2\leq C$. Define
\begin{equation}
\label{theta1}
\theta_1 = 1 - \frac{2}{C\left( \lambda_{max} \|\mathbf{R}_n\|_2 + \|\mathbf{R}\|_2 \right)} \quad \textrm{~and~} \quad \theta_0 = (1-\theta_1)\cdot C \|\mathbf{R}\|_2 + 1.
\end{equation}
Note that $1=\theta_0 - (1-\theta_1)\cdot C \|\mathbf{R}\|_2$ and $\theta_1<1$, we obtain
\begin{align*}
\|\mathbf{X}^T \mathbf{R}\mathbf{X}-\mathbf{X}\|_2 &\leq \|\mathbf{X}^T \mathbf{R}\mathbf{X}\|_2 + \|\mathbf{X}\|_2 =  \|\mathbf{X}^T \mathbf{R}\mathbf{X}\|_2 + \left( \theta_0 - (1-\theta_1)\cdot C \|\mathbf{R}\|_2 \right) \|\mathbf{X}\|_2 \nonumber \\ & \leq \|\mathbf{X}^T \mathbf{R}\mathbf{X}\|_2 + \theta_0 \|\mathbf{X}\|_2 - (1-\theta_1)\cdot \|\mathbf{X}\|_2 \cdot \|\mathbf{R}\|_2 \cdot \|\mathbf{X}\|_2 \nonumber \\ & \leq \|\mathbf{X}^T \mathbf{R}\mathbf{X}\|_2 + \theta_0 \|\mathbf{X}\|_2 - (1-\theta_1)\cdot  \|\mathbf{X}^T \mathbf{R}\mathbf{X}\|_2 \nonumber \\ & = \theta_0 \|\mathbf{X}\|_2 + \theta_1 \|\mathbf{X}^T \mathbf{R} \mathbf{X}\|_2,
\label{inequlity1}
\end{align*}
which implies that
\begin{align*}
\|\mathbf{X}^T \mathbf{R}\mathbf{X}\|_2 &\leq \|\mathbf{X}^T \mathbf{R} \mathbf{X}-\mathbf{X}\|_2 + \|\mathbf{X}\|_2 \leq \theta_0 \|\mathbf{X}\|_2 +  \theta_1 \|\mathbf{X}^T \mathbf{R} \mathbf{X}\|_2 + \|\mathbf{X}\|_2 = \\ &= (1+\theta_0) \|\mathbf{X}\|_2 +  \theta_1 \|\mathbf{X}^T \mathbf{R} \mathbf{X}\|_2.
\end{align*}
Hence, we obtain
\begin{equation}
\|\mathbf{X}^T \mathbf{R} \mathbf{X}\|_2 \leq \frac{1+\theta_0}{1-\theta_1}\|\mathbf{X}\|_2.
\label{result1}
\end{equation}

Furthermore, define (note that $\mathbf{R}$ is nonsingular by the assumption)
\begin{equation}
\label{theta2}
\theta_2 = -1 -\lambda_{max} \frac{\|\mathbf{R}_n\|_2}{\|\mathbf{R}\|_2}.
\end{equation}
By definitions of $\theta_0$, $\theta_1$ and $\theta_2$ in~\eqref{theta1} and~\eqref{theta2}, respectively, we have
\begin{equation*}
(1-\theta_2) C \|\mathbf{R}_n\|_2 + \frac{(1+\theta_0)(1+\theta_2)}{(1-\theta_1)\lambda_{max}} = 0,
\end{equation*}
which implies that 
\begin{align*}
(1-\theta_2)& \|\mathbf{X}^T \mathbf{R}_n\mathbf{X}\|_2 + \frac{(1+\theta_0)(1+\theta_2)}{(1-\theta_1)\lambda_{max}} \|\mathbf{X}\|_2 \\ & \leq (1-\theta_2) C \|\mathbf{X}\|_2 \|\mathbf{R}_n\|_2 + \frac{(1+\theta_0)(1+\theta_2)}{(1-\theta_1)\lambda_{max}} \|\mathbf{X}\|_2 \\ & = \left\{ (1-\theta_2) C \|\mathbf{R}_n\|_2 + \frac{(1+\theta_0)(1+\theta_2)}{(1-\theta_1)\lambda_{max}} \right\} \|\mathbf{X}\|_2 = 0.
\end{align*}
Then, using the following inequalities
\begin{align*}
\|\mathbf{X}^T \mathbf{R}_n \mathbf{X}-\mathbf{X}\|_2 &\leq \|\mathbf{X}^T \mathbf{R}_n\mathbf{X}\|_2 + \|\mathbf{X}\|_2 = \theta_2 \|\mathbf{X}^T \mathbf{R}_n\mathbf{X}\|_2 + (1-\theta_2)\|\mathbf{X}^T \mathbf{R}_n\mathbf{X}\|_2 \\ & \qquad + \left( 1- \frac{(1+\theta_0)(1+\theta_2)}{(1-\theta_1)\lambda_{max}} \right) \|\mathbf{X}\|_2 + \frac{(1+\theta_0)(1+\theta_2)}{(1-\theta_1)\lambda_{max}} \|\mathbf{X}\|_2 \\ & \leq \left( 1- \frac{(1+\theta_0)(1+\theta_2)}{(1-\theta_1)\lambda_{max}} \right) \|\mathbf{X}\|_2 +  \theta_2 \|\mathbf{X}^T \mathbf{R}_n\mathbf{X}\|_2 \\ & \qquad + \left\{ (1-\theta_2) C \|\mathbf{R}_n\|_2 + \frac{(1+\theta_0)(1+\theta_2)}{(1-\theta_1)\lambda_{max}} \right\} \|\mathbf{X}\|_2 \\ & = \left( 1- \frac{(1+\theta_0)(1+\theta_2)}{(1-\theta_1)\lambda_{max}} \right) \|\mathbf{X}\|_2 +  \theta_2 \|\mathbf{X}^T \mathbf{R}_n\mathbf{X}\|_2,
\end{align*}
we deduce that $\|\mathbf{X}^T \mathbf{R}_n \mathbf{X}-\mathbf{X}\|_2 \leq \theta_3 \|\mathbf{X}\|_2 +  \theta_2 \|\mathbf{X}^T \mathbf{R}_n\mathbf{X}\|_2$, where $\theta_3 = \left( 1- \frac{(1+\theta_0)(1+\theta_2)}{(1-\theta_1)\lambda_{max}} \right)$. This implies 
\begin{align*}
\|\mathbf{X}^T \mathbf{R}_n\mathbf{X}\|_2 &\geq  \|\mathbf{X}\|_2 -\|\mathbf{X}^T \mathbf{R}_n \mathbf{X}-\mathbf{X}\|_2  \geq \|\mathbf{X}\|_2 - \theta_3 \|\mathbf{X}\|_2 -  \theta_2 \|\mathbf{X}^T \mathbf{R}_n\mathbf{X}\|_2 =\\ &= (1-\theta_3) \|\mathbf{X}\|_2 - \theta_2 \|\mathbf{X}^T \mathbf{R}_n\mathbf{X}\|_2,
\end{align*}
which leads to the inequality
\begin{equation}
\|\mathbf{X}^T \mathbf{R}_n\mathbf{X}\|_2 \geq \frac{1-\theta_3}{1+\theta_2} \|\mathbf{X}\|_2.
\label{result2}
\end{equation}

Since $\frac{1-\theta_3}{1+\theta_2} = \frac{1+\theta_0}{(1-\theta_1)\lambda_{max}}>0$, combine~\eqref{result1} and~\eqref{result2} to get 
\begin{equation*}
\mathcal{I}_{\mathbf{R}, \mathbf{R}_n} (\mathbf{X}) = \frac{\|\mathbf{X}^T \mathbf{R} \mathbf{X}\|_2}{\|\mathbf{X}^T \mathbf{R}_n \mathbf{X}\|_2}  \leq \frac{\frac{1+\theta_0}{1-\theta_1}}{\frac{1-\theta_3}{1+\theta_2}} = \lambda_{max},
\end{equation*}
which yields the boundedness of $\mathcal{I}_{\mathbf{R}, \mathbf{R}_n}$.

Finally, let us show that transformations $\mathbf{X}^{opt}_{i}$ ($i=1,2$) in \eqref{trans1}-\eqref{trans2} are the required solutions. Denote by $I_K$ a $K$-dimensional identity matrix. By noting that $\mathbf{\Phi}_n$ and $\mathbf{\Phi}_{adj}$ are unitary matrices, we obtain
\begin{align*}
& \mathcal{I}_{\mathbf{R}, \mathbf{R}_n} (\mathbf{X}^{opt}_2) = \frac{\|(\mathbf{X}^{opt}_2)^T \mathbf{R} \mathbf{X}^{opt}_2\|_2}{\|(\mathbf{X}^{opt}_2)^T  \mathbf{R}_n \mathbf{X}^{opt}_2\|_2} = \frac{\|\mathbf{\Phi}^T_{adj} \mathbf{\Lambda}^{-1/2}_n \mathbf{\Phi}^T_n \mathbf{R} \mathbf{\Phi}_n \mathbf{\Lambda}^{-1/2}_n \mathbf{\Phi}_{adj}\|_2}{\|\mathbf{\Phi}^T_{adj} \mathbf{\Lambda}^{-1/2}_n \mathbf{\Phi}^T_n \mathbf{R}_n \mathbf{\Phi}_n \mathbf{\Lambda}^{-1/2}_n \mathbf{\Phi}_{adj}\|_2} \\ &= \frac{\|\mathbf{\Lambda}^{-1/2}_n \mathbf{\Phi}^T_n \mathbf{R} \mathbf{\Phi}_n \mathbf{\Lambda}^{-1/2}_n \|_2}{\|\mathbf{\Lambda}^{-1/2}_n \mathbf{\Phi}^T_n \mathbf{R}_n \mathbf{\Phi}_n \mathbf{\Lambda}^{-1/2}_n \|_2} \left( = \frac{\|\mathbf{W}^T_n \mathbf{R} \mathbf{W}_n\|_2}{\|\mathbf{W}^T_n \mathbf{R}_n \mathbf{W}_n\|_2} = \mathcal{I}_{\mathbf{R}, \mathbf{R}_n} (\mathbf{X}^{opt}_1)\right) = \frac{\|\mathbf{\Lambda}_{adj}\|_2}{\|I_K\|_2} =\lambda_{max},
\end{align*}
which yields the required result and we prove the theorem completely.
\end{pf}

\section*{Appendix 2: Proof of Theorem 2}\label{thm:consistent}
\begin{pf}
By condition \eqref{ConsistentCondition1} and the boundedness of $\|\mathbf{\Lambda}^{-1/2}_n\|$ we have
\begin{align*}
\|\widehat{\mathbf{\Phi}}_n \widehat{\mathbf{\Lambda}}^{-1/2}_n - \mathbf{\Phi}_n \mathbf{\Lambda}^{-1/2}_n\| & \leq \|\widehat{\mathbf{\Phi}}_n \widehat{\mathbf{\Lambda}}^{-1/2}_n - \widehat{\mathbf{\Phi}}_n \mathbf{\Lambda}^{-1/2}_n \| + \| \widehat{\mathbf{\Phi}}_n \mathbf{\Lambda}^{-1/2}_n -  \mathbf{\Phi}_n \mathbf{\Lambda}^{-1/2}_n\| \leq \\ & \|\widehat{\mathbf{\Phi}}_n \| \cdot \| \widehat{\mathbf{\Lambda}}^{-1/2}_n - \mathbf{\Lambda}^{-1/2}_n \| + \| \widehat{\mathbf{\Phi}}_n -  \mathbf{\Phi}_n \| \cdot \|\mathbf{\Lambda}^{-1/2}_n\| \leq (1+C) \omega
\end{align*}
which implies that $\|\widehat{\mathbf{W}}_n - \mathbf{W}_n\| = \|\widehat{\mathbf{\Phi}}_n \widehat{\mathbf{\Lambda}}^{-1/2}_n - \mathbf{\Phi}_n \mathbf{\Lambda}^{-1/2}_n\| \leq (1+C) \omega$. 

Similarly, it is easy to obtain the following inequality 
\begin{equation*}
\|\widehat{\mathbf{\Lambda}}^{-1/2}_n\| \leq \|\widehat{\mathbf{\Lambda}}^{-1/2}_n-\mathbf{\Lambda}^{-1/2}_n\| + \|\mathbf{\Lambda}^{-1/2}_n\| \leq \omega + C.
\end{equation*}

Then, using the above inequalities and definitions of $\mathbf{R}_{adj}$ and $\widehat{\mathbf{R}}_{adj}$, we conclude that
\begin{align*}
\|\widehat{\mathbf{R}}_{adj}-\mathbf{R}_{adj}\| & = \|\widehat{\mathbf{W}}^T_n \widehat{\mathbf{R}} \widehat{\mathbf{W}}_n- \mathbf{W}^T_n \mathbf{R} \mathbf{W}_n\| \leq \\ & \|\widehat{\mathbf{W}}^T_n \widehat{\mathbf{R}} \widehat{\mathbf{W}}_n- \widehat{\mathbf{W}}^T_n \widehat{\mathbf{R}} \mathbf{W}_n\|  + \| \widehat{\mathbf{W}}^T_n \widehat{\mathbf{R}} \mathbf{W}_n - \widehat{\mathbf{W}}^T_n \mathbf{R} \mathbf{W}_n \| + \| \widehat{\mathbf{W}}^T_n \mathbf{R} \mathbf{W}_n - \mathbf{W}^T_n \mathbf{R} \mathbf{W}_n\| \leq \\ & \|\widehat{\mathbf{W}}^T_n \widehat{\mathbf{R}} \| \cdot \| \widehat{\mathbf{W}}_n- \mathbf{W}_n \| + \|\widehat{\mathbf{W}}^T_n\| \cdot \| \widehat{\mathbf{R}} - \mathbf{R}\| \cdot  \|\mathbf{W}_n \| +  \|\widehat{\mathbf{W}}^T_n - \mathbf{W}^T_n \| \cdot \|\mathbf{R} \mathbf{W}_n \| = \\ & \left( \|\widehat{\mathbf{W}}^T_n \widehat{\mathbf{R}} \| + \| \mathbf{R} \mathbf{W}_n\| \right) \| \widehat{\mathbf{W}}_n- \mathbf{W}_n \| + \| \widehat{\mathbf{W}}_n\| \cdot \| \mathbf{W}_n \| \cdot \| \widehat{\mathbf{R}} - \mathbf{R}\| \leq \\ & \left( \|(\widehat{\mathbf{\Phi}}_n \widehat{\mathbf{\Lambda}}^{-1/2}_n)^T \widehat{\mathbf{R}} \| + \| \mathbf{R} (\mathbf{\Phi}_n \mathbf{\Lambda}^{-1/2}_n)\| \right) (1+C) \omega + \|\widehat{\mathbf{\Phi}}_n \widehat{\mathbf{\Lambda}}^{-1/2}_n \| \cdot \|\mathbf{\Phi}_n \mathbf{\Lambda}^{-1/2}_n \|  \omega \leq
\end{align*}

\begin{align*}
& \left( \|\widehat{\mathbf{\Phi}}_n\| \cdot \|\widehat{\mathbf{\Lambda}}^{-1/2}_n \| \cdot \| \mathbf{R} \| + \| \mathbf{R} \| \cdot \|\mathbf{\Phi}_n \| \cdot\|\mathbf{\Lambda}^{-1/2}_n\| \right) (1+C) \omega  \\ & \qquad\qquad\qquad\qquad\qquad + \|\widehat{\mathbf{\Phi}}_n\| \cdot \|\widehat{\mathbf{\Lambda}}^{-1/2}_n \| \cdot \|\mathbf{\Phi}_n \| \cdot \| \mathbf{\Lambda}^{-1/2}_n \| \cdot \omega \leq \\ & \left( (\omega + C) \| \mathbf{R} \|  + \| \mathbf{R} \| C \right) (1+C) \omega + (\omega + C) C \omega = \\ &  \left( 2 C (1+C) \| \mathbf{R} \| + C^2 \right) \omega + \left( (1+C) \| \mathbf{R} \| +C \right) \omega^2 = O (\omega),
\end{align*}
which yields the required result.
\end{pf}

%\end{comment}

%% HERE WE DECLARE THE BIBLIOGRAPHYSTYLE TO USE AND THE BIBLIOGRAPHY DATABASE
\bibliographystyle{ECA_jasa}
\bibliography{proposal_refs_2}

\end{document}